\documentclass[11pt,a4paper]{article}
\usepackage{xy,amsmath,amssymb,slashed,url,bm,textgreek,upgreek}
\usepackage[table]{xcolor}
\usepackage{bbm,array,amsfonts,wrapfig,lscape,float,mathtools,multirow,longtable}
\usepackage{graphicx}
\usepackage{epstopdf}
 
\def\tt{{\mathfrak t}}

\def\be{\begin{equation}}
\def\ee{\end{equation}}

\def\hat{\widehat}

\def\D{{\mathcal D}}

\def\R{{\mathbb R}}
\def\C{{\mathbb C}}

\def\D{{\mathcal D}}
\def\[{\bigl [}

\def\]{\bigr ]}

\def\Z{{\mathbb Z}}

\def\bar{\overline}

\font\teneurm=eurm10 \font\seveneurm=eurm7  \font\fiveeurm=eurm5
\newfam\eurmfam
\textfont\eurmfam=\teneurm \scriptfont\eurmfam=\seveneurm
\scriptscriptfont\eurmfam=\fiveeurm

\font\teneusm=eusm10 \font\seveneusm=eusm7 \font\fiveeusm=eusm5
\newfam\eusmfam
\textfont\eusmfam=\teneusm \scriptfont\eusmfam=\seveneusm
\scriptscriptfont\eusmfam=\fiveeusm

\font\tencmmib=cmmib10 \skewchar\tencmmib='177
\font\sevencmmib=cmmib7 \skewchar\sevencmmib='177
\font\fivecmmib=cmmib5 \skewchar\fivecmmib='177
\newfam\cmmibfam
\textfont\cmmibfam=\tencmmib \scriptfont\cmmibfam=\sevencmmib
\scriptscriptfont\cmmibfam=\fivecmmib

\title{Generalized Symmetries in 2D from String Theory: SymTFTs, Intrinsic Relativeness, and Anomalies of Non-invertible Symmetries}

 \author[a,b,c]{Sebasti\'an Franco,}
\author[d]{Xingyang Yu}
\affiliation[a]{Physics Department, The City College of the CUNY\\
	160 Convent Avenue, New York, NY 10031, USA}
\affiliation[b]{Physics Program and \textsuperscript{$c$}Initiative for the Theoretical Sciences\\
	The Graduate School and University Center, The City University of New York\\
	365 Fifth Avenue, New York NY 10016, USA}
\affiliation[d]{Physics Department, Robeson Hall, Virginia Tech, Blacksburg, VA 24061, USA}

\abstract{Generalized global symmetries, in particular non-invertible and categorical symmetries, have become a focal point in the recent study of quantum field theory (QFT). In this paper, we investigate aspects of symmetry topological field theories (SymTFTs) and anomalies of non-invertible symmetries for 2D QFTs from a string theory perspective. Our primary focus is on an infinite class of 2D QFTs engineered on D1-branes probing toric Calabi-Yau 4-fold singularities. We derive 3D SymTFTs from the topological sector of IIB supergravity and discuss the resulting 2D QFTs, which can be intrinsically relative or absolute. For intrinsically relative QFTs, we propose a sufficient condition for them to exist. For absolute QFTs, we show that they exhibit non-invertible symmetries with an elegant brane origin. Furthermore, we find that these non-invertible symmetries can suffer from anomalies, which we discuss from a top-down perspective. Explicit examples are provided, including theories for $Y^{(p,k)}(\mathbb{P}^2)$, $Y^{(2,0)}(\mathbb{P}^1\times \mathbb{P}^1)$, and $\C^4/\Z_4$ geometries.}
\begin{document}
\maketitle
\flushbottom

\section{Introduction}
The study of generalized global symmetries \cite{Gaiotto:2014kfa} has been instrumental in advancing our understanding of quantum field theory (QFT) and string theory. Notably, generalized global symmetries, including \emph{non-invertible symmetries}, are ubiquitous in 2D QFTs, garnering significant attention over the past several years (see, for instance, \cite{Verlinde:1988sn, Fuchs:2002cm, Frohlich:2004ef, Bhardwaj:2017xup, Chang:2018iay, Thorngren:2019iar, Komargodski:2020mxz, Thorngren:2021yso}). Non-invertible symmetries are described by the mathematical notion of (higher) \emph{fusion categories} (see, e.g., \cite{Etingof:2002vpd}), thus they are sometimes also referred to as \emph{categorical symmetries}.

A useful approach for investigating categorical symmetries is to extend the 2D spacetime manifold into a 3D bulk, where a topological field theory (TFT) lives. This methods works due to the topological nature of generalized global symmetries, which are symmetries captured by topological operators in QFTs \cite{Gaiotto:2014kfa}. Such topological information is encoded in the 3D TFT with two boundaries. One of the boundaries is physical, containing all local information of the 2D theory one is interested in, whereas the other boundary is topological and determines boundary conditions. This construction allows the decoupling of the global structure of a 2D QFT from its complicated local information, providing a clean approach to generalized global symmetries. The 3D bulk TFT is now commonly referred to as a \emph{symmetry TFT (SymTFT)} (see, e.g., \cite{Reshetikhin:1991tc, Turaev:1992hq, Barrett:1993ab, Witten:1998wy, Fuchs:2002cm, Kirillov:2010nh, Kapustin:2010if, Kitaev:2011dxc, Fuchs:2012dt, Freed:2012bs}).\footnote{See also \cite{Freed:2018cec, Freed:2022qnc, Kaidi:2022cpf, Antinucci:2022vyk, Bhardwaj:2023wzd, Bhardwaj:2023ayw, Baume:2023kkf, bhardwaj:2023bbf, Brennan:2024fgj, Heckman:2024oot, Antinucci:2024zjp, Argurio:2024oym, Bonetti:2024cjk, Apruzzi:2024htg, Bhardwaj:2024qrf, DelZotto:2024tae, GarciaEtxebarria:2024fuk} for a partial list of more recent work.} 

In many cases, the 3D bulk is auxiliary. This is because it is possible to move the topological boundary, merge it with the physical boundary, and thereby achieve a genuine 2D system. However, there are cases where the 3D TFT does not admit any topological boundary condition, so it is not possible to collapse the 3D bulk into a 2D system. The 3D TFT is now intrinsic, instead of auxiliary, and the corresponding 2D theory living on the physical boundary is known as a \emph{relative QFT} \cite{Freed:2012bs}. In many works in literature, the term ``relative QFT'' is frequently also employed to indicate the physical boundary theory of the SymTFT with a topological boundary. To avoid ambiguity, in this paper we introduce the notion of \emph{intrinsically relative QFT} to denote a theory whose bulk TFT does not admit any topological boundary condition. In other words, the bulk TFT is \emph{intrinsic} instead of auxiliary. 

When the SymTFT bulk admits a topological boundary, the 2D system obtained via shrinking the bulk is called an \emph{absolute QFT}. If a SymTFT admits multiple boundary conditions, it will have multiple absolute QFT descendants, which are connected via topological manipulations, e.g., finite gauging and stacking local counterterms/SPT phases. This leads to a powerful aspect of the SymTFT, namely it can capture the anomalies for non-invertible symmetries. Consider an absolute QFT associated with a certain topological boundary condition for the SymTFT with a non-invertible symmetry. An attempt to gauge this symmetry would amount to changing the topological boundary condition. If the new boundary condition is not allowed by the SymTFT, then we end up with an obstruction to gauging, which corresponds to a 't Hooft anomaly for the non-invertible symmetry (see, e.g., \cite{Zhang:2023wlu, Kaidi:2023maf, Cordova:2023bja, Antinucci:2023ezl}).

A powerful perspective for studying QFTs and their symmetries involves embedding them into string theory. Several works have been devoted to building generalized global symmetries from a top-down perspective in QFTs that admit a string theory realization via geometry engineering or brane probes. There are two primary questions in this approach: how to build topological symmetry operators and how to derive the SymTFTs. For a QFT engineered at a conical singularity $Y$, topological symmetry operators originate from either branes wrapping cycles ``at infinity" along $\partial Y$ \cite{Apruzzi:2022rei, GarciaEtxebarria:2022vzq, Heckman:2022muc} (see also \cite{Heckman:2022xgu, Apruzzi:2023uma, Bah:2023ymy, Dierigl:2023jdp, Cvetic:2023plv, Acharya:2023bth}), or geometric fibers degenerated ``at infinity" \cite{Lawrie:2023tdz}. The non-topological dynamics of these asymptotic boundary objects (branes or fibers) is decoupled from the QFT localized at the conical singularity, but its topological effects remain. The SymTFT for the QFT, on the other hand, is obtained from the dimensional reduction of the topological sector of 10D/11D supergravity actions along the asymptotic boundary $\partial Y$ \cite{Apruzzi:2021nmk} (see also \cite{vanBeest:2022fss, Apruzzi:2022rei, Lawrie:2023tdz, Apruzzi:2023uma, Baume:2023kkf, Yu:2023nyn, Basile:2023zng, Apruzzi:2024htg, Heckman:2024oot, DelZotto:2024tae, Braeger:2024jcj}).

Despite extensive work from this top-down perspective, the focus has predominantly been on QFTs in spacetime dimensions greater than 2 ($D>2$). Furthermore, to our knowledge, a string theory approach for anomalies of non-invertible symmetries has yet to be explored in the literature. The purpose of this paper is to provide a systematic study of aspects of SymTFTs and anomalies of non-invertible symmetries in 2D QFTs from a string theory point of view. The main setup we will work on is an infinite class of 2D QFTs engineered on D1-branes probing toric Calabi-Yau 4-fold singularities. This class of 2D QFTs enjoys a nice intersecting brane construction, known as the \emph{brane brick model} \cite{Franco:2015tna,Franco:2015tya,Franco:2016nwv,Franco:2016qxh}. This stringy setup, which we will review in Section 4, provides a geometric way to investigate many aspects of 2D gauge theories. In this work we will see how the geometry gives rise to categorical symmetry structure of these 2D QFTs.
\subsection*{Organization of the paper}

We start in Section 2 by reviewing some known aspects of SymTFTs and the defect group. We start with intermediate defect groups and their polarizations (pairs), and then discuss how this information is nicely described by SymTFTs and their topological boundary conditions. We discuss how the intrinsic relativeness of a QFT is extracted from the absence of a topological boundary condition of a SymTFT, equivalently from the absence of a Lagrangian subgroup of the defect group. For relative QFTs with absolute QFT descendants, we discuss how the gauging and anomalies of the finite symmetries are captured from the SymTFT and the defect group's point of view. In particular, we discuss how the anomalies of non-invertible symmetries can be reformulated in terms of the defect group language, in the case when non-invertible symmetries are derived from gauging invertible ones. 

In Section 3, we give a lightning review of the main stage of this paper: 2D QFTs engineered from Calabi-Yau 4-folds probed by D1-branes.  This is an infinite family of theories that enjoys a quiver gauge theory description, as well as a Type IIA intersecting brane realization known as \emph{brane brick models}.

Section 4 presents a systematic investigation of aspects of categorical symmetries for these 2D QFTs from IIB string theory. We start with an explicit derivation of 3D SymTFTs from the topological sector of the IIB supergravity, as well as intermediate defect groups from the geometric data. By investigating the topological boundary conditions, as well as the possible polarization choices, we find that the resulting 2D QFTs can be intrinsically relative and propose a sufficient condition for it. For cases of 2D absolute QFTs, we obtain two general classes of polarizations, which enjoy invertible and non-invertible symmetries, respectively, connected by finite gauging. The anomalies of the (non)-invertible symmetries are then captured by obstruction to certain topological boundary conditions. From the brane perspective, we show how the non-invertible symmetry line operators can be explicitly computed from dimensionally reducing the brane worldvolume action. The polarizations and anomalies for finite symmetries are then translated from the possible brane configurations, which are allowed to end ``at infinity".

In Section 5, we illustrate our ideas in explicit examples. We first consider an infinite class of theories for the $Y^{(p,k)}(\mathbb{P}^2)$ geometry, and discuss a sufficient condition for the associated 2D QFTs to be intrinsically relative theories. We then consider $Y^{(2,0)}(\mathbb{P}^1\times \mathbb{P}^1)$, whose associated 2D QFTs admit many polarizations, one of which enjoys a non-anomalous non-invertible symmetry Rep$(D_4)$. The final example we present is $\C^4/\Z_4$, for which we show that the corresponding 2D QFT enjoys an anomalous non-invertible symmetry.

\section{SymTFTs, Defect Groups, and Anomalies}

This work focuses on the categorical finite symmetries of 2D QFTs associated with toric Calabi-Yau 4-folds in Type IIB string theory. The main methodologies employed are the defect group and the Symmetry TFT (SymTFT). In this section, we provide a concise overview of these two notions, with a particular emphasis on how they capture essential information regarding a QFT, such as its defect group, relativeness, and anomalies of (non-invertible) finite symmetries.

Let us start with a short review of the intermediate defect group in even-dimensional QFTs following \cite{Lawrie:2023tdz}. For QFTs in $2k$-dimensional spacetime, associated with self-dual gauge fields, there are ``light'' excitations and ``heavy'' defects, both with $(k-1)$-dimensional worldvolumes in spacetime. The charges associated with light objects reside in the lattice $\Lambda$, while those of heavy objects are within the dual lattice $\Lambda^*$, which is a refinement of $\Lambda$ known as $\mathbb{Q}$-refinement \cite{Deser:1997se}. The intermediate defect group\footnote{The ``intermediate'' here is named after the fact that the flux operators associated with the $(k-1)$-dimensional defects belong to the intermediate cohomology class $H^k(M_{2k},\mathbb{D})$.}, denoted as $\mathbb{D}$ and defined as
\begin{equation}
    \mathbb{D}=\Lambda^*/\Lambda,
\end{equation}
quantifies the non-integer nature of the Dirac pairing among $(k-1)$-dimensional charged defects. The quotient in the above equation is interpreted as equivalence classes under screening of dynamical objects \cite{THOOFT19781}. At this level, a $2k$-dimensional QFT lacks an inherent scalar-valued partition function. Instead, it possesses a partition vector, characteristic of a \emph{relative} QFT \cite{Freed:2012bs}. 

Establishing a consistent quantum field theory with a well-defined partition function on any closed spacetime manifold necessitates selecting a sublattice for the $(k-1)$-dimensional objects, ensuring the Dirac pairing be integral. This selection corresponds to picking a Lagrangian subgroup $L\subset \mathbb{D}$, often termed as choosing a \emph{polarization} \cite{Belov:2006jd} (see also, e.g., \cite{Freed:2012bs, Gukov:2020btk, Lawrie:2023tdz}). Given a choice of $L$, the resulting absolute QFT enjoys a $(k-1)$-form global symmetry, given by $L^\vee=\mathbb{D}/L$, which can be embedded in the following exact sequence
\begin{equation}\label{eq: defect group sequence}
    1\rightarrow L \rightarrow \mathbb{D} \rightarrow L^{\vee} \rightarrow 1.
\end{equation}

The results of picking polarizations can, in general, be classified into three cases.
\begin{itemize}
    \item \textbf{Case 1: Intrinsically Relative QFT.} There is no Lagrangian subgroup of $\mathbb{D}$. This, in general, happens in $2k=4s+2=2, 6, 10, \cdots$ dimensions, where the defect group $\mathbb{D}$ is equipped with a symmetric Dirac pairing, i.e., there exists non-trivial self-pairing for defects. In this case, the associated QFT does not allow a well-defined partition function on a $2k$-dimensional spacetime manifold but only captures a partition vector. As a result, this QFT should be defined as the boundary theory of a $(2k+1)$-dimensional topological field theory (TFT), where the partition vector is the boundary state of the Hilbert space for the TFT \cite{Freed:2012bs}. From now on, we will refer to this type of relative QFTs as \emph{intrinsically relative} to distinguish them from those admitting polarizations.\footnote{In this context, due to the fact that all defects are, strictly speaking, topological operators in $(2k+1)$-dimensional TFT, there is no clear notion of distinguishing them as charged objects or symmetry operators. Therefore, it is not clear how to rigorously define global symmetries associated with the defect group for the relative QFT.}
    \item \textbf{Case 2: Absolute QFT with non-anomalous symmetry.} There is a Lagrangian subgroup $L\subset \mathbb{D}$, and the exact sequence in (\ref{eq: defect group sequence}) splits. The $(k-1)$-form global symmetry $L^\vee$ can then be uplifted as a Lagrangian subgroup $L^\vee \cong \overline{L}$ back into the defect group $\mathbb{D}$ so that $\mathbb{D}=L\oplus \overline{L}$ \cite{Gukov:2020btk}.  This means the $(k-1)$-form global symmetry $L^\vee$ is non-anomalous. Gauging this symmetry leads to another absolute QFT, whose associated polarization is given by $\overline{L}$. This amounts to regarding an absolute QFT as not just one Lagrangian subgroup $L$, but a pair of Lagrangian subgroups $(L,\overline{L})$, known as a \emph{polarization pair} \cite{Lawrie:2023tdz}. The gauging is then simply the flip manipulation $(L,\overline{L})\rightarrow (\pm \overline{L}, L)$, where ``$+$'' and ``$-$'' correspond to the case $2k=4s$ and $2k=4s+2$, respectively\footnote{There are more involved cases that a pair of Lagrangian subgroups $(L, \overline{L})$ still do not fully specify the global structure of a QFT, but only up to generalized charge conjugations. In those cases, a polarization pair is defined by a generator of $L$ and a generator of $\overline{L}$. We refer the reader to \cite{Lawrie:2023tdz} for more details.}. Namely, the gauged theory is associated with the polarization $\overline{L}$ with global symmetry $\bar{L}^\vee \cong L$. Notably, there are possibly multiple uplifts of $L^\vee$ for a given $L$. This encodes the absolute QFTs with the same polarization $L$ but is distinguished by stacking symmetry-protected topological (SPT) phases/local counterterms, which also naturally encodes the discrete torsion choices for the gauging.
    \item \textbf{Case 3: Absolute QFT with anomalous symmetry.} There is a Lagrangian subgroup $L\subset \mathbb{D}$, but the exact sequence in (\ref{eq: defect group sequence}) does not split. In this case, $L^\vee$ cannot be embedded back in the defect group $\mathbb{D}$ as a subgroup. According to the discussion in Case 2, this means one cannot gauge $L^\vee$ symmetry to arrive at another polarization. This obstruction to gauging captures a 't Hooft anomaly for the $(k-1)$-form symmetry $L^\vee$.
\end{itemize}
In principle, only in the above Case 1, a $2k$-dimensional TFT is necessarily introduced. However, we will see below that even for relative QFTs admitting polarizations, it is also useful to introduce a one-dimensional higher TFT to capture the information of global symmetries and their anomalies. The associated TFT is known as a symmetry TFT (SymTFT) \cite{Reshetikhin:1991tc, Turaev:1992hq, Barrett:1993ab, Witten:1998wy, Fuchs:2002cm, Kirillov:2010nh, Kapustin:2010if, Kitaev:2011dxc, Fuchs:2012dt, Freed:2012bs, Freed:2018cec, Freed:2022qnc, Kaidi:2022cpf, Baume:2023kkf, Yu:2023nyn}, which will be the main focus of the rest of this section.

\subsection{SymTFTs and Gapped Boundary Conditions}

\begin{figure}[h]
    \centering
    \includegraphics[width=11cm]{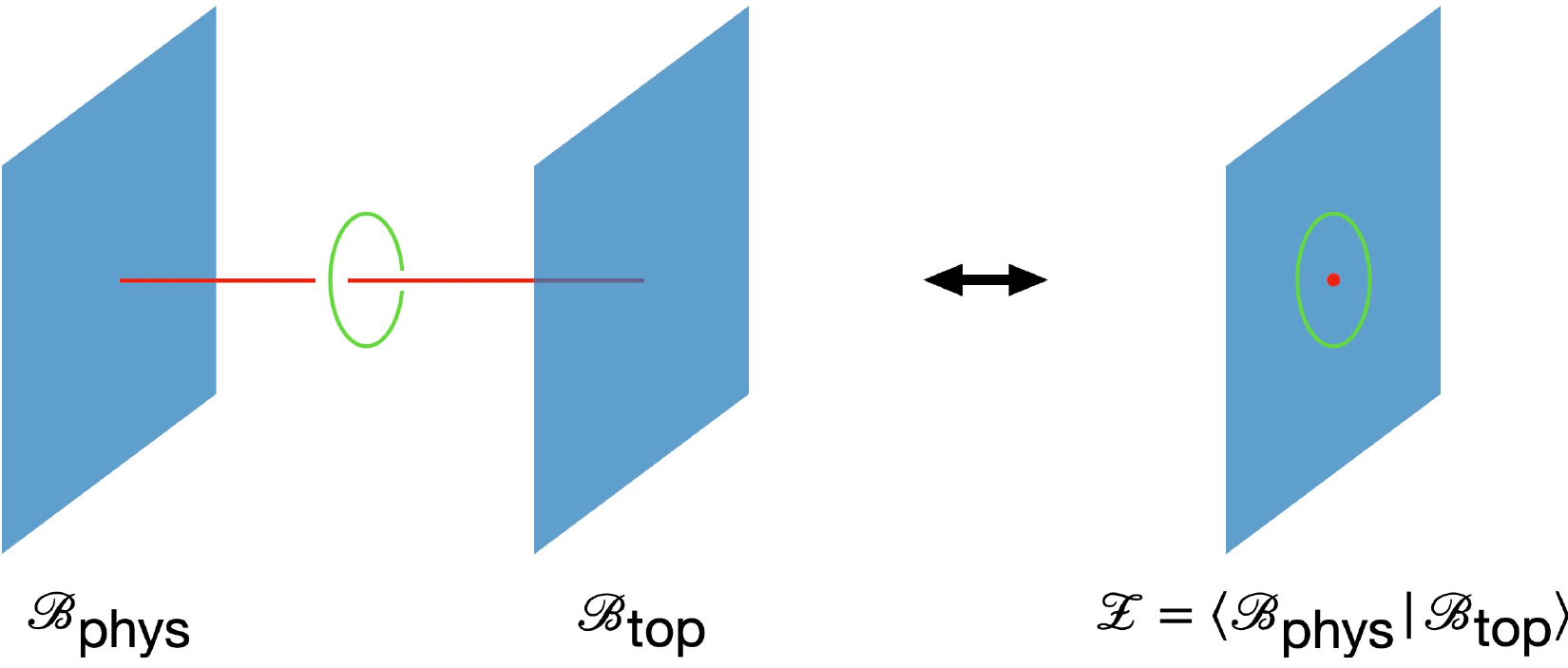}
    \caption{Absolute QFT as a SymTFT slab. The red line denotes an operator terminating on the gapped boundary $\mathcal{B}_{\text{top}}$, and the green line denotes an operator linking with it. After shrinking the slab, the resulting theory has a well-defined partition function $\mathcal{Z}$. The green loop is the symmetry operator measuring the charge of the heavy defect labeled by the red point.}
    \label{fig: symtft slab}
\end{figure}

A SymTFT is the (D+1)-dimensional TFT capturing the generalized global symmetries of a D-dimensional QFT \cite{Reshetikhin:1991tc, Turaev:1992hq, Barrett:1993ab, Witten:1998wy, Fuchs:2002cm, Kirillov:2010nh, Kapustin:2010if, Kitaev:2011dxc, Fuchs:2012dt, Freed:2012bs, Freed:2018cec, Freed:2022qnc, Kaidi:2022cpf, Baume:2023kkf, Yu:2023nyn}. In many cases, the SymTFT is defined on a bulk manifold $M_{D+1}$ with two boundaries. One boundary is physical, encoding the local information of the D-dimensional QFT we are interested in. This boundary QFT is exactly the relative QFT as we discussed before, intuitively speaking ``relative'' to a (D+1)-dimensional bulk \cite{Freed:2012bs}. The other boundary is topological, imposing gapped boundary conditions for fields in the SymTFT, which specifies the global structure of the QFT. Under the TFT quantization, there are boundary states in the Hilbert space associated with the physical boundary $\langle \mathcal{B}_{\text{phys}}|$ and the topological boundary $|\mathcal{B}_{\text{top}}\rangle$. The partition function of the resulting D-dimensional QFT is then given by the inner product
\begin{equation}\label{eq: absolute QFT partition function}
    \mathcal{Z}_{\mathcal{B}_{\text{phys}}, \mathcal{B}_{\text{top}}}=\langle \mathcal{B}_{\text{phys}}|\mathcal{B}_{\text{top}}\rangle.
\end{equation}
This can be understood as shrinking the SymTFT bulk slab and ending up with a well-defined D-dimensional absolute QFT. See Figure \ref{fig: symtft slab} for an illustration. This corresponds to picking a polarization for the relative QFT \cite{Belov:2006jd, Freed:2012bs, Gukov:2020btk, Lawrie:2023tdz}.

From the operator algebra perspective, picking gapped boundary conditions corresponds to picking the \emph{Lagrangian subalgebra}\footnote{A Lagrangian subalgebra $\mathcal{L}\subset \mathcal{A}$ is a maximally isotropic subspace with respect to the natural scalar product in $\mathcal{A}$.} generated by a subset of operators in the SymTFT \cite{Kapustin:2010hk}. Generally speaking, TFT fields can pick the Dirichlet condition or the Neumann (i.e., free) condition at the gapped boundary. This will enable some topological operators to terminate on the topological boundary (red line in Figure \ref{fig: symtft slab}) while others continue along the topological boundary (green loop in Figure \ref{fig: symtft slab} after pulling into the bulk). On the one hand, those continuing along the topological boundary are topological operators that generate the global symmetry for the absolute QFT. On the other hand, the Lagrangian subalgebra is generated by operators terminating on the gapped boundary with trivial linking and thus can be condensed simultaneously. 

An illustrative example is the SymTFT for 4D $\mathfrak{su}(2)$  $\mathcal{N}=4$ super Yang-Mills (SYM) theory \cite{Witten:1998wy},
\begin{equation}
    S_5=\frac{2\pi}{2}\int_{M_5}b_2 \cup \delta c_2, 
\end{equation}
where $b_2$ and $c_2$ are $\mathbb{Z}_2$-valued cochains, under the defect group $\mathbb{D}=\mathbb{Z}_2^{(e)}\times \mathbb{Z}_2^{(m)}$, respectively. The spectrum of surface topological operators for this 5D TFT reads
\begin{equation}
    U_{(m,n)}(M_2)=\exp\left(\pi i \int_{M_2}mb_2+nc_2 \right),
\end{equation}
where $m,n\in \{0,1\}$. 
Using the TFT quantization
\begin{equation}
    \left[ b_{ij}(x), c_{kl}(y) \right]=\frac{i}{\pi}\epsilon_{ijkl} \delta(x-y),
\end{equation}
one obtains the non-trivial linking generated by
\begin{equation}
    U_{(1,0)}(M_2)U_{(0,1)}(M_2')=U_{(0,1)}(M_2')U_{(1,0)}(M_2)e^ {\pi i~\text{Link}(M_2, M_2')}.
\end{equation}

On the physical boundary, local information (i.e., local operators and their correlation functions) of the SYM gives rise to the relative QFT with Lie algebra $\mathfrak{su}(2)$. The global form of the gauge group, however, is captured by the gapped boundary conditions on the topological boundary. For example, the gauge group $SU(2)$ is associated with the following boundary condition
\begin{equation}
    b_2~\text{Dirichelt}, c_2~\text{Neumann}.
\end{equation}
The Lagrangian subalgebra corresponding to this boundary condition is generated by 
\begin{equation}
  \{ U_{(0,0)}, U_{(1,0)} \},
\end{equation}
within which all operators have trivial linking, thus can be simultaneously condensed. This aligns with the fact that $U_{(0,0)}$ and $U_{(1,0)}$ can terminate on the gapped boundary because of the Dirichlet condition for $b_2$.

\subsubsection*{Lagrangian subgroups and polarization pairs}

From the defect group perspective, a Lagrangian subalgebra associated with the gapped boundary conditions can be alternatively derived from a Lagrangian subgroup of the defect group: $L\subset \mathbb{D}$. The integral Dirac pairing of the sublattice required by a QFT with a well-defined partition function corresponds to the trivial linking between topological operators, which builds a Lagrangian subalgebra of the TFT.

The correspondence between the defect group and the SymTFT is discussed in detail in, e.g., \cite{Lawrie:2023tdz, Gukov:2020btk}. Briefly speaking, given an intermediate defect group in a $2k$-dimensional QFT with a generic form $\mathbb{D}=\oplus_i \Z_{N_i}$ and its corresponding Dirac pairing, one can write down a SymTFT with quadratic single-derivative action (namely Chern-Simons or BF-type terms),
\begin{equation}\label{eq: symtft for dirac pairing}
    S_{\text{symTFT}}[b_i]=\int_{M_{2k+1}} \frac{1}{2} b_i Q_{ij}\delta b_j,
\end{equation}
where $b_i$ is the background gauge field for $\Z_{N_i}\subset \mathbb{D}$, and the matrix $Q_{ij}$ is the coefficient of the $\mathbb{Q}/\mathbb{Z}$-valued bilinear Dirac pairing on the defect group $\mathbb{D}$ \cite{Lawrie:2023tdz, Gukov:2020btk}. A Lagrangian subgroup of the defect group with integral Dirac pairing is then translated into a gapped boundary condition of the SymTFT. This boundary condition then specifies a subset of $k$-dimensional topological operators in the SymTFT with trivial linking relations, generating a Lagrangian subalgebra.

Now, back to the 4D $\mathfrak{su}(2)$ SYM example, one can easily see the defect group is given by 
\begin{equation}
    \mathbb{D}=\mathbb{Z}_2^{(e)}\times \mathbb{Z}_2^{(m)},
\end{equation}
whose background fields are $b_2$ and $c_2$, respectively. The defect group is equipped with an anti-symmetric Dirac pairing
\begin{equation}\label{eq: Dirac pairing for su(2)}
    \begin{pmatrix}
    0 & \frac{1}{2} \\
    -\frac{1}{2} & 0
    \end{pmatrix}.
\end{equation}
Denote elements in the defect group as $(a,b)\in \mathbb{D}$, where $a\in \mathbb{Z}_2^{(e)}$ and $b\in \mathbb{Z}_2^{(m)}$. The Lagrangian subgroup for the global form $SU(2)$ is given by 
\begin{equation}\label{eq: Lagrangian subgroup for SU(2)}
    L_{SU(2)}=\mathbb{Z}_2^{(m)},
\end{equation}
whose generating element $(0,1)$ enjoys integer Dirac pairing under (\ref{eq: Dirac pairing for su(2)}). The 1-form global symmetry $G^{(1)}_{SU(2)}$ for the resulting absolute $SU(2)$ theory is then given by condensing the symmetry operators for $\mathbb{Z}_2^{(m)}$, namely treating it as the ``gauged'' part via the quotient
\begin{equation}
    G^{(1)}_{SU(2)}=L^\vee_{SU(2)} \equiv \mathbb{D}/\mathbb{Z}_2^{(m)}
\end{equation}

We remark that picking a gapped boundary condition/Lagrangian subalgebra may not fully fix the global structure of an absolute QFT. For example, after choosing the global form of the gauge group $SU(2)$, one can decide whether to further stack a $\mathbb{Z}_2$ symmetry-protected topological (SPT) phase $\exp \left(\frac{i\pi}{2}\int_{M_4}\mathcal{P}(b_2) \right)$ to the theory, with $\mathcal{P}(b_2)$ the Pontryagin square of $b_2$. Following the notation in \cite{Kaidi:2022uux}, we denote the theory with and without SPT stacking as $SU(2)_0$ and $SU(2)_1$, respectively. These two theories, associated with the same Lagrangian subgroup, translate in two possible embeddings of $L^{\vee}$ in $\mathbb{D}$ as another Lagrangian subgroup $\overline{L}$
\begin{equation}
\begin{split}
    &L^\vee_{SU(2)}\rightarrow \overline{L}=\mathbb{Z}_2^{(e)},\\
    \text{or}~ &L^\vee_{SU(2)}\rightarrow \overline{L}=\mathbb{Z}_2^{(d)}
\end{split}
\end{equation}
so that 
\begin{equation}
    L\oplus \overline{L}=\mathbb{D},
\end{equation}
where $\mathbb{Z}_2^{(d)}$ is the diagonal $\mathbb{Z}_2$ subgroup of $\mathbb{D}$.

The fully specified global structure of the theory is then given by a pair of Lagrangian subgroups $(L,\overline{L})$ 
\begin{equation}\label{eq: polar pairs for SU(2)}
\begin{split}
    &SU(2)_0: (L=\mathbb{Z}_2^{(m)}, \overline{L}=\mathbb{Z}_2^{(e)}),\\
    &SU(2)_1: (L=\mathbb{Z}_2^{(m)}, \overline{L}=\mathbb{Z}_2^{(d)}),
\end{split}
\end{equation}
As we discussed previously in this section, this is a refined notion of polarization known as the polarization pair \cite{Lawrie:2023tdz}.

\subsection{Relativeness: SymTFTs without Any Gapped Boundary Condition}

It is not always possible to pick gapped boundary conditions for a given SymTFT. Alternatively speaking, SymTFTs exist whose spectrum of operators does not allow any Lagrangian subalgebra. In this case, one can only introduce a physical boundary for the SymTFT. The relative QFT living on the physical boundary does not admit any polarization/absolute QFT descendants. For the sake of clarity, we will refer to this type of QFT as \emph{intrinsically relative}.  Instead of a well-defined partition function in (\ref{eq: absolute QFT partition function}), the best one can do for this type of theory is to define a partition vector $|\mathcal{B}_{\text{phys}}\rangle$ in the Hilbert space from the quantization of the corresponding SymTFT. 

An illustrative example is a $(4s+3)$-dimensional SymTFT such as a $U(1)_2$ Chern-Simons theory
\begin{equation}\label{eq: 4l+3 CS theory}
    S_{4s+3}=\frac{2}{4\pi}\int_{M_{4s+3}}c_{2s+1}\wedge d c_{2s+1},
\end{equation}
where $c_{2s+1}$ is a  $(2s+1)$-form $U(1)$ field.  The topological operators of this theory are 
\begin{equation}\label{eq: Wilson lines in 4k+3 CS}
    U_{m}(M_{2s+1})=\exp \left(i m \int_{M_{2s+1}}c_{2s+1} \right), m\in \mathbb{Z}_2
\end{equation}
which, in the case of $s=0$, are just Wilson loops for 3D Chern-Simons Theory. Since we are just working in the case of Chern-Simons level $k=2$, there is only one non-trivial operator $U_1(M_{2s+1})$ with self-linking (see, e.g., \cite{Gaiotto:2014kfa})
\begin{equation}
U_1(M_{2s+1})U_1(M_{2s+1}')=U_1(M_{2s+1}')U_1(M_{2s+1})e^{\pi i~\text{Link}(M_{2s+1},M_{2s+1}')}.
\end{equation}
Therefore, there is no subset of operators with trivial linking, i.e., no Lagrangian subalgebra. This implies that there is no well-defined gapped boundary conditions for $c_{2s+1}$. This means on the physical boundary, the $(4s+2)$-dimensional QFT does not admit any polarization/absolute QFT descendent. In other words, its relativeness to the bulk SymTFT is inevitable.

\subsubsection*{No Lagrangian subgroup}

This relativeness can also be seen from the perspective of the defect group. The defect group for the $(4s+2)$-dimensional QFT whose SymTFT is $U(1)_2$ Chern-Simons theory is given by a single factor 
\begin{equation}
    \mathbb{D}=\mathbb{Z}_2
\end{equation}
with the symmetric Dirac pairing matrix, which in this case is just a $1\times 1$ matrix with a single element 
\begin{equation}
    \frac{1}{2}.
\end{equation}

As discussed in the 4D $\mathfrak{su}(2)$ SYM case (\ref{eq: Lagrangian subgroup for SU(2)}), one attempts to find a Lagrangian subgroup of the defect group associated with a possible well-defined gapped boundary condition. However, one necessary condition for a Lagrangian subgroup is
\begin{equation}\label{eq: necessary condition for L}
    |L|^2=|\mathbb{D}|,
\end{equation}
with $|G|$ the order of the group $G$.\footnote{For more details of Lagrangian subgroups and their refined treatment via Heisenberg groups, we refer the reader to \cite{Witten:1998wy, Tachikawa:2013hya, Gukov:2020btk} for physical discussions and \cite{mumford2006tata} for a mathematical review.} Therefore, $\mathbb{Z}_2$ has no Lagrangian subgroup; thus, this SymTFT does not admit a gapped boundary condition. That is to say, the associated $(4s+2)$ QFT is intrinsically relative. 

When $s=0$, (\ref{eq: 4l+3 CS theory}) is  a 3D $U(1)_2$ Chern-Simons theory. One possible associated relative QFT is a 2D chiral CFT without modular-invariant partition function but with conformal blocks as the partition vector, e.g., the chiral WZW model. When $s=1$, (\ref{eq: 4l+3 CS theory}) is equivalent to a 7D Chern-Simons theory for 3-form gauge field $c_3$. One possible class of the associated intrinsically relative QFTs includes 6D superconformal field theories (SCFTs) with the defect group $\mathbb{Z}_2$, e.g., $A_1$ $\mathcal{N}=(2,0)$ SCFT.

\subsection{Anomalies: Obstructions to Certain Gapped Boundary Conditions}

Generally speaking, there are two notions for defining the 't Hooft anomalies of a global symmetry. One is defined as the obstruction of gauging the symmetry, while the other is defined as the obstruction to a symmetry-protected trivially gapped phase. For invertible symmetries, these two notions coincide, while for non-invertible symmetries, they can be non-equivalent \cite{Choi:2023xjw}. In this work, we will use the former notion and translate the gauging obstruction as the obstruction to certain gapped boundary conditions for SymTFTs \cite{Kaidi:2023maf, Antinucci:2023ezl, Zhang:2023wlu, Cordova:2023bja}.

\subsubsection{Anomalous Invertible Symmetries}

For invertible global symmetries, anomalies are nicely captured by $(D+1)$-dimensional invertible TFTs via the anomaly inflow construction \cite{Callan:1984sa}. The resulting invertible TFTs can then be embedded into SymTFTs by promoting all fields to be dynamical. For example, an anomalous $\mathbb{Z}_2$ symmetry with background field $A_1$ in 2D QFT can be described by a 3D term
\begin{equation}\label{eq: anomaly inflow}
    \pi  \int_{M_3}\frac{1}{2}A_1\cup \delta A_1,
\end{equation}

\subsubsection*{$\mathbb{Z}_2$ twisted Dijkgraaf-Witten theory}

 The anomaly term (\ref{eq: anomaly inflow}) can be embedded into a 3D SymTFT as a Dijkgraaf-Witten theory \cite{Dijkgraaf:1989pz} by promoting $A_1$ into a dynamical field $a_1$ with a  $H^3(U(1),\mathbb{Z}_2)=\mathbb{Z}_2$ twist,
\begin{equation}\label{eq: 3D twisted Z_2 DW theory}
    S_3=\frac{2\pi}{2}\int_{M_3}a_1\cup \delta \hat{a}_1+\frac{2\pi}{4}\int_{M_3}a_1\cup \delta a_1.
\end{equation}
Picking the gapped boundary condition 
\begin{equation}\label{eq: e condition for twisted Z2 DW}
    a_1~\text{Dirichlet}, ~\hat{a}_1~\text{Neumann}
\end{equation}
leads to a 2D absolute QFT with $\mathbb{Z}_2$ symmetry generated by $\exp\left( \pi i \int_{M_1}\hat{a}_1 \right)$ with anomalies (\ref{eq: anomaly inflow}). This ``electric'' boundary condition is allowed, since $\exp\left( \pi i \int_{M_1}a_1 \right)$ has trivial self-linking such that
\begin{equation}
    \{1, \exp\left(\pi i \int_{M_1}a_1 \right) \}
\end{equation}
generates a Lagrangian subalgebra.

An attempt to gauging the $\mathbb{Z}_2$ symmetry amounts to condensing the $\mathbb{Z}_2$ symmetry operator $\exp\left( \pi i \int_{M_1}\hat{a}_1 \right)$, leading to a naive ``magnetic'' boundary condition 
\begin{equation}\label{eq: m condition for twisted Z2 DW}
    a_1~\text{Neumann},~\hat{a}_1~\text{Dirichlet}.
\end{equation}
However, unlike the line operator for $a$, the topological line $\exp\left( \pi i \int_{M_1}\hat{a}_1 \right)$ has a non-trivial self-linking (see, e.g., \cite{Levin:2012yb})
\begin{equation}
    \langle \exp\left( \pi i \int_{M_1}\hat{a}_1 \right)\exp\left( \pi i \int_{M_1'}\hat{a}_1 \right) \rangle \sim e^{\pi i~\text{Link}(M_1,M_1')}.
\end{equation}
This means it cannot generate a Lagrangian subalgebra and be condensed. Therefore, the ``magnetic'' boundary condition (\ref{eq: m condition for twisted Z2 DW}) is obstructed, capturing anomalies for the $\mathbb{Z}_2$ symmetry.

\subsubsection*{$U(1)_4$ Chern-Simons theory}

In addition to the Dijkgraaf-Witten theory (\ref{eq: 3D twisted Z_2 DW theory}), another SymTFT able to realize the anomaly (\ref{eq: anomaly inflow}) is similar to the previous relativeness example (\ref{eq: 4l+3 CS theory}), but this time a $U(1)_4$ Chern-Simons theory,
\begin{equation}
    S_3=\frac{4}{4\pi}\int_{M_3}c_1\wedge  dc_1,
\end{equation}
where $c_1$ is a 1-form $U(1)$ gauge field.
The line operator spectrum of this theory is given by Wilson lines (\ref{eq: Wilson lines in 4k+3 CS}) with $m\in \mathbb{Z}_4$. The essential difference from the $U(1)_2$ case in (\ref{eq: 4l+3 CS theory}) is now the operator $U_2(M_1)$ has trivial self-linking. Therefore, it is possible to pick a Lagrangian subalgebra generated by 
\begin{equation}\label{eq: Lagrangian subalgebra of U(1)4}
    \{1, U_2(M_1)\}
\end{equation}
and define an associated gapped boundary condition
\begin{equation}\label{eq: boundary condition of U(1)4}
    c_1|_{\partial M_3}=A_1,
\end{equation}
with $A_1$ a $\mathbb{Z}_2$-valued gauge field. This boundary condition leads to a 2D absolute QFT with a $\mathbb{Z}_2$ symmetry, whose anomaly is captured by (\ref{eq: anomaly inflow}). 

This anomaly can be easily seen from the operator algebra perspective. Notice that (\ref{eq: Lagrangian subalgebra of U(1)4}) is the only possible Lagrangian subalgebra because all other line operators have non-trivial linkings. This obstruction to other boundary conditions by condensing topological operators implies the anomaly.

\subsubsection*{Anomaly as an obstruction to polarization pair}

In the defect group language, the anomalies of $\mathbb{Z}_2$ symmetry can be interpreted as the obstruction to a polarization pair. Let us briefly discuss how this works for both the Dijkgraaf-Witten theory and the Chern-Simons theory as SymTFTs.

For the twisted Dijkgraaf-Witten theory (\ref{eq: 3D twisted Z_2 DW theory}), the associated 2D relative QFT has the defect group 
\begin{equation}
    \mathbb{D}=\mathbb{Z}_2^{(a)}\times \mathbb{Z}_2^{(\hat{a})}
\end{equation}
with a symmetric Dirac pairing matrix
\begin{equation}
    \begin{pmatrix}
        \frac{1}{2}& \frac{1}{2}\\
        \frac{1}{2}& 0
    \end{pmatrix}.
\end{equation}
The Lagrangian subgroup corresponding to the gapped boundary condition (\ref{eq: e condition for twisted Z2 DW}) is given by 
\begin{equation}
    L=\mathbb{Z}_{2}^{(\hat{a})}\subset \mathbb{D}
\end{equation}
generated by $(0,1)\in \mathbb{D}$, due to the integer Dirac pairing 
\begin{equation}
    (0,1)\begin{pmatrix}
        \frac{1}{2}& \frac{1}{2}\\
        \frac{1}{2}& 0
    \end{pmatrix}\begin{pmatrix}
        0\\
        1
    \end{pmatrix}\in \mathbb{Z}
\end{equation}
It is straightforward to check all other non-trivial subgroups of $\mathbb{D}$ do not have integer Dirac pairing; thus, the global symmetry 
\begin{equation}
    L^{\vee}=\mathbb{D}/\mathbb{Z}_2^{(\hat{a})}
\end{equation}
cannot be uplifted to any Lagrangian subgroup. This obstructs the polarization associated with $L=\mathbb{Z}_2^{\hat{a}}$ to be promoted into a polarization pair, which implies the global symmetry $L^{\vee}$ of the absolute QFT is anomalous. 

For the $U(1)_4$ Chern-Simons theory, the associated 2D relative QFT has the defect group 
\begin{equation}
    \mathbb{D}=\mathbb{Z}_4
\end{equation}
with a single $1\times 1$ matrix element $\frac{1}{4}$ as the symmetric Dirac pairing. The Lagrangian subgroup corresponding to the gapped boundary condition (\ref{eq: boundary condition of U(1)4}) is given by 
\begin{equation}
    L=\mathbb{Z}_2\subset \mathbb{D},
\end{equation}
due to the integer Dirac pairing $\frac{1}{4}\times 2^2\in \mathbb{Z}$. The global symmetry $\mathbb{Z}_2=L^{\vee}\equiv \mathbb{D}/\mathbb{Z}_2$ of the resulting absolute theory cannot be embedded back in $\mathbb{D}=\mathbb{Z}_4$ as a Lagrangian subgroup. This is due to the fact that the short exact sequence
\begin{equation}
    1\rightarrow L=\Z_2 \rightarrow \Z_4 \rightarrow L^\vee=\Z_2 \rightarrow 1
\end{equation}
does not split. This obstructs the polarization associated with $L=\mathbb{Z}_2$ to be promoted as a polarization pair, which implies the anomaly of the $\mathbb{Z}_2$ symmetry.

\subsubsection{Anomalous Non-invertible Symmetries}

The anomalies of non-invertible symmetries can be captured following the same philosophy as those of invertible symmetries. Namely, starting with a polarization/gapped boundary condition, referred to as ``electric'', whose corresponding absolute QFT enjoys a non-invertible symmetry, the obstruction to its ``magnetic'' boundary condition via simultaneously condensing topological operators implies the anomaly for the non-invertible symmetry \cite{Kaidi:2023maf, Antinucci:2023ezl, Cordova:2023bja}.

An illustrative example is discussed in \cite{Kaidi:2023maf}, where the authors constructed an anomalous non-invertible symmetry given by the $\mathbb{Z}_2\times \mathbb{Z}_2$ Tambara-Yamagami (TY) fusion category \cite{TAMBARA1998692}. Here, we briefly review their result. Consider a 2D QFT enjoying a $\mathbb{Z}_2^{(a)}\times \mathbb{Z}_2^{(b)}\times \mathbb{Z}_2^{(c)}$ symmetry with anomalies captured by a 3D invertible TFT
\begin{equation}\label{eq: anomaly TFT for non-inver1}
    \pi \int_{M_3}\frac{1}{2}A_1\cup \delta A_1+A_1\cup B_1\cup C_1,
\end{equation}
where $A_1, B_1$ and $C_1$ are $\Z_2$ cochains as background fields for $\Z_2^{(a)}, \Z_2^{(b)}$ and $\Z_2^{(c)}$, respectively. Gauging the $\Z_2^{(b)}\times \Z_2^{(c)}$ symmetry, the topological line operator $L(M_1; a)$ for $\Z_2^{(a)}$ then fails to be gauge-invariant, but can be cured by stacking a 1D $\Z_2\times \Z_2$ Dijkgraaf-Witten theory \cite{Kaidi:2023maf, Yu:2023nyn}
\begin{equation}\label{eq: noninvertible line for Z2Z2 TY}
    \mathcal{N}(M_1;b_1,c_1)\equiv L(M_1;b_1,c_1)\int \D\hat{\phi}_0 \mathcal{D}\phi_0 \exp \left( \pi i \int_{M_1}\hat{\phi}_0\cup \delta \phi_0+\phi_0 \cup b_1-\hat{\phi}_0 \cup c_1 \right),
\end{equation}
where $\phi_0$ and $\hat{\phi}_0$ $\mathbb{Z}_2$-valued 0-cochains and $b_1, c_1$ are dynamical fields from promoting $B_1, C_1$ via gauging. This non-invertible defect line generates a $\Z_2\times \Z_2$ Tambara-Yamagami (TY) fusion categorical symmetry \cite{TAMBARA1998692}, which we will denote as TY$(\Z_2\times \Z_2)$. 

Note that this categorical symmetry is inherited from the invertible symmetry $\Z_2^{a}$. Its anomaly can then be regarded as also inherited from the self-anomaly given by the first term in (\ref{eq: anomaly TFT for non-inver1}). More specifically, one can consider embedding this anomaly TFT into a SymTFT
\begin{equation}\label{eq: symtft for anomalous noninver}
    S_3=\frac{2\pi}{2}\int_{M_3}a_1\cup \delta \hat{a}_1+b_1\cup \delta \hat{b}_1+c_1\cup \delta \hat{c}_1+\frac{1}{2}a_1\cup \delta a_1+a_1\cup b_1\cup c_1.
\end{equation}
The absolute QFT enjoying the TY$(\Z_2\times \Z_2)$ categorical symmetry is given by the following ``electric'' gapped boundary condition
\begin{equation}
    a_1, \hat{b}_1,\hat{c}_1~\text{Dirichlet},~\hat{a}_1, b_1, c_1~\text{Neumann}
\end{equation}
under which the topological line operator 
\begin{equation}
    L_{\hat{a}_1,b_1,c_1}(M_1)=\int\D\hat{\phi}_0 \D\phi_0 \exp \left( \pi i \int_{M_1} \hat{a}_1 \right)  \exp \left( \pi i \int_{M_1} \hat{\phi}_0\cup \delta \phi_0+\phi_0 \cup b_1-\hat{\phi}_0 \cup c_1 \right)
\end{equation}
corresponds to the symmetry generator $\mathcal{N}(M_1;b_1,c_1)$ (\ref{eq: noninvertible line for Z2Z2 TY}). It is computed in \cite{Kaidi:2023maf} that there is always a non-trivial linking for $L_{\hat{a}_1,b_1,c_1}(M_1)$ 
\begin{equation}\label{eq: linking of non-invertible lines in KNZZ}
    \langle L_{\hat{a}_1,b_1,c_1}(M_1)L_{\hat{a}_1,b_1,c_1}(M_1')  \rangle\sim e^{ \pi i~\text{Link}(M_1,M_1')}\neq 1,
\end{equation}
due to the term $\frac{1}{2}a_1\cup \delta a_1$ in the SymTFT. This non-trivial linking prevents the operator $L_{\hat{a}_1,b_1,c_1}(M_1)$ from forming any Lagrangian subalgebra, implying the obstruction to the ``magnetic'' gapped boundary condition on which the $L_{\hat{a}_1,b_1,c_1}(M_1)$ line can terminate.

\subsubsection*{Anomaly from obstruction to polarization pair}

Let us try understanding this anomaly for non-invertible symmetries from the defect group and the polarization pair perspective. The single-derivative terms in SymTFT (\ref{eq: symtft for anomalous noninver}) show that the associated 2D relative QFT has the defect group 
\begin{equation}
   \mathbb{D}= \Z_2^{(a)}\times \Z_2^{(\hat{a})}\times \Z_2^{(b)}\times \Z_2^{(\hat{b})}\times \Z_2^{(c)}\times \Z_2^{(\hat{c})},
\end{equation}
with the Dirac pairing matrix
\begin{equation}\label{eq: dirac pairing for anomalous noninver}
    \begin{pmatrix}
        \textcolor{red}{\frac{1}{2}}& \frac{1}{2}& 0 & 0 & 0 & 0\\
        \frac{1}{2}& 0 & 0 & 0 & 0 & 0\\
        0& 0 & 0 & \frac{1}{2} & 0 & 0\\
        0& 0 & \frac{1}{2} & 0 & 0 & 0\\
        0& 0 & 0 & 0 & 0 & \frac{1}{2}\\
        0& 0 & 0 & 0 & \frac{1}{2} & 0\\
    \end{pmatrix}.
\end{equation}
This defect group admits many Lagrangian subgroups with integral Dirac pairing under (\ref{eq: dirac pairing for anomalous noninver}). One of them is $L_1=\Z_2^{(\hat{a})}\times \Z_2^{(\hat{b})} \times \Z_2^{(\hat{c})}$, corresponding to an absolute 2D QFT with global symmetry,
\begin{equation}
    L^\vee_1=\mathbb{D}/L_1\cong \Z_2^{(a)}\times \Z_2^{(b)}\times \Z_2^{(c)},
\end{equation}
with the anomaly TFT as shown previously in (\ref{eq: anomaly TFT for non-inver1}). One might try promoting this polarization $L_1$ into a polarization pair, but will then find out the $\Z_2^{(a)}\times \Z_2^{(b)}\times \Z_2^{(c)}$ does not enjoy a integral Dirac pairing under (\ref{eq: dirac pairing for anomalous noninver}). This means $L_1^\vee$ cannot be embedded into $\mathbb{D}$ as a Lagrangian subgroup, which implies the obstruction to gauging the full $L_1^\vee$ symmetry.

However, it is possible to gauge part of the $L_1^\vee$, namely the $\Z_2^{(b)}\times \Z_2^{(c)}$ symmetry.\footnote{More precisely, we mean to gauge the $\Z_2^{(b)}\times \Z_2^{(c)}$ without discrete torsion.} The resulting polarization 
is given by flipping the role of $b,c$ with $\hat{b},\hat{c}$, resulting in the Lagrangian subgroup $L_2=\Z_2^{(\hat{a})}\times \Z_2^{(b)}\times \Z_2^{(c)}$. The associated quotient $L_2^\vee$ of the defect group reads
\begin{equation}
    L_2^\vee=\mathbb{D}/L_2\cong \Z_2^{(a)}\times \Z_2^{(\hat{b})}\times \Z_{2}^{(\hat{c})}.
\end{equation}
However, in the presence of the cubic term in $(\ref{eq: symtft for anomalous noninver})$, $L^\vee_2$ is not the genuine global symmetry for the absolute QFT, but it is promoted to a non-invertible symmetry.
\begin{equation}\label{eq: anomaly TFT for non-inver1}
    L^\vee_2 \rightarrow G=\text{TY}(\Z_2\times \Z_2)
\end{equation}
This promotion can be understood as replacing the direct product in $L^\vee_2$ between $\Z_2^{(a)}$ and $\Z_2^{(\hat{b})}\times \Z_2^{(\hat{c})}$ with a $\Z_2^{(a)}$ extension of the invertible fusion category Vec$(\Z_2^{(\hat{b})}\times \Z_2^{(\hat{c})})$.\footnote{We use the notation Vec$(G)$=Rep$(\C[G]^*)$ as the category of G-graded vector spaces.}\footnote{In general, TY$(G)$ fusion category ($G$ is a finite group) can be derived from a $\Z_2$ extension of the Vec$(G)$ fusion category. Physically speaking, this implies the theory is self-dual under gauging the $G$ symmetry. See, e.g., \cite{Choi:2023xjw, Perez-Lona:2023djo, Diatlyk:2023fwf, Choi:2021kmx} for more discussion on this $\Z_2$ extension.} The result of this non-trivial group extension is exactly the TY$(\Z_2\times \Z_2)$ non-invertible symmetry.

Now it is also straightforward to see why this non-invertible symmetry is anomalous. An attempt to gauge this non-invertible symmetry translates in promoting $L_2^\vee$ into a Lagrangian subgroup in order to build a polarization pair refined from the $L_2$ polarization. However, one can check that $\Z_2^{(a)}\times \Z_2^{(\hat{b})}\times \Z_{2}^{(\hat{c})}$ does not enjoy a integral Dirac pairing, due to the $\frac{1}{2}$ element labeled in red in the pairing matrix (\ref{eq: dirac pairing for anomalous noninver}). Therefore, $L_2^\vee$ cannot be uplifted to a Lagrangian subgroup of $\mathbb{D}$, implying the anomaly which the descendent global symmetry $G=$TY$(\Z_2\times \Z_2)$ suffers from.\footnote{So far there is no universal construction for polarization pairs of QFTs with non-invertible symmetries. We hope to come back to this problem in the future, and strongly welcome other researchers to do so.} Note that the $\frac{1}{2}$ in red in (\ref{eq: dirac pairing for anomalous noninver}) is exactly the coefficient responsible for the self-anomaly term $\frac{1}{2} A_1\cup \delta A_1$ in (\ref{eq: anomaly TFT for non-inver1}), which we have argued from the SymTFT gapped boundary condition point of view, is the origin of the anomaly for the non-invertible symmetry.

\section{2D QFTs from Calabi-Yau 4-Folds}

In this paper, we will focus on 2D $(0,2)$ gauge theories engineered on D1-branes probing toric Calabi-Yau 4-folds.\footnote{The amount of SUSY may be enhanced depending on the geometry.} These theories can be efficiently encoded in terms of {\it brane brick models}, which are obtained from D1-branes at Calabi-Yau 4-fold singularities by T-duality. We refer the reader to \cite{Franco:2015tna,Franco:2015tya,Franco:2016nwv,Franco:2016qxh} for detailed discussions.

\begin{figure}[h]
	\centering
	\includegraphics[height=5cm]{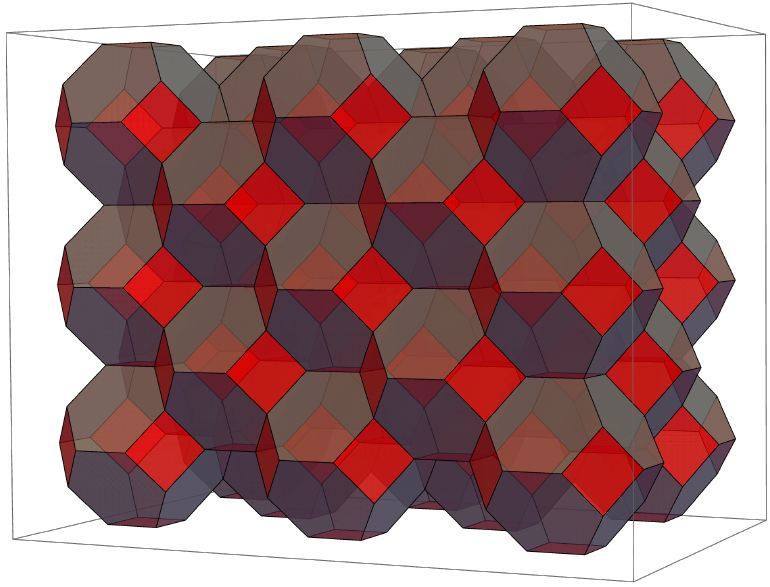}
\caption{Brane brick model for $\mathbb{C}^4$. Here we show multiple copies of the unit cell.}
	\label{BBM_C4}
\end{figure}

A brane brick model is a Type IIA brane configuration consisting of D4-branes wrapping a 3-torus $\mathbb{T}^3$ 
and suspended from an NS5-brane that wraps a holomorphic surface $\Sigma$ intersecting with $\mathbb{T}^3$. The holomorphic surface $\Sigma$ is the zero locus of the Newton polynomial defined by the toric diagram of the Calabi-Yau 4-fold. As an example, Figure \ref{BBM_C4} shows the brane brick model for $\mathbb{C}^4$. The corresponding QFT is the $(8,8)$ theory obtained from dimensionally reducing 4D $\mathcal{N}=4$ SYM to 2D.

Table \ref{Brane brick-config} summarizes the basic ingredients in a Brane Brick Model. The $(246)$ directions are compactified on a 3-torus, while the 2D gauge theory lives on the $(01)$ directions common to all the branes. 

\begin{table}[ht!!]
\centering
\begin{tabular}{l|cccccccccc}
\; & 0 & 1 & 2 & 3 & 4 & 5 & 6 & 7 & 8 & 9 \\
\hline
$\text{D4}$ & $\times$ & $\times$ & $\times$ & $\cdot$ & $\times$ & $\cdot$ & $\times$ & $\cdot$ & $\cdot$ & $\cdot$  \\
$\text{NS5}$ & $\times$ & $\times$ & \multicolumn{6}{c}{----------- \ $\Sigma$ \ ------------} & $\cdot$ & $\cdot$ \\
\end{tabular}
\caption{Brane brick model configuration.}
\label{Brane brick-config}
\end{table}

Brane brick models, or equivalently their dual periodic quivers, fully specify the 2D $(0,2)$ quiver gauge theories on the worldvolume of D1-branes probing toric Calabi-Yau 4-folds. The dictionary connecting brane brick models to the corresponding gauge theories is summarized in Table \ref{tbrick}. 

\begin{table}[h]
\centering
\begin{tabular}{|c|c|c|}
\hline
{\bf Brane Brick Model} \ \ &  {\bf  Gauge Theory} \ \ \ \ \ \ \  & {\bf Periodic Quiver} \ \ \ 
\\
\hline\hline
Brick  & $U(N)$ Gauge group & Node \\
\hline
Oriented face  & Bifundamental chiral field from & Oriented (black) arrow 
\\
between bricks $i$ and $j$ & node $i$ to node $j$ (adjoint if $i=j$) & from node $i$ to node $j$ \\
\hline
Unoriented square face  & Bifundamental Fermi field between & Unoriented (red) line \\
between bricks $i$ and $j$ & nodes $i$ and $j$ (adjoint if $i=j$) & between nodes $i$ and $j$  \\
\hline
Edge  & $J$- or $E$-term & Plaquette encoding \\ 
& & a $J$- or an $E$-term \\
\hline
\end{tabular}
\caption{Dictionary between brane brick models and $2d$ $(0,2)$ gauge theories.}
\label{tbrick}
\end{table}

Brane brick models streamline the connection between gauge theory and the underlying geometry (see, e.g., \cite{Franco:2015tya,Franco:2016qxh,Franco:2016fxm,Franco:2018qsc}). Additional results on brane brick models can be found in  \cite{Franco:2017cjj,Franco:2021elb,Franco:2023tyf}.

Before moving to the general discussion of categorical symmetries of the brane brick models, we remark that in general, the gauge groups for the brane brick model can be $\prod_{i}U(N+M_i)$, with $M_i$ the number of fractional branes on $i$-th quiver node. In the context of finite categorical symmetries in this work, the concrete form of the gauge groups would not matter much. However, their diagonal $U(1)$ center may lead to continuous $U(1)$ 1-form symmetries, which may have interesting interplay (e.g., mixed anomalies or 2-groups \cite{Benini:2018reh}) with the non-invertible finite symmetries. We leave this for a future work. 

\section{Aspects of Generalized Symmetries in 2D from String Theory}

Having reviewed how 2D QFTs can be engineered from Calabi-Yau 4-folds in Type IIB string theory, in this section, we discuss how to extract categorical symmetries for this infinite class of theories from the geometric data. In particular, we will present a general prescription for deriving 3D SymTFTs via dimensional reduction of the topological sector of type IIB supergravity. Furthermore, we also discuss how the topological operators in the 3D SymTFTs can be derived from the brane worldvolume actions via the construction of branes ``at infinity''. The gapped boundary conditions are translated into boundary conditions for various IIB fluxes ``at infinity'', under which the dimensionally reduced brane worldvolume actions serve as charged objects generating the Lagrangian subalgebra and symmetry operators for the associated absolute 2D QFTs.

\subsection{3D Symmetry TFTs from Type IIB string theory}

Consider 2D QFTs associated with singular Calabi-Yau 4-folds $Y$, whose base spaces are Sasaki-Einstein 7-manifold, probed by D1-branes. The IIB string theory background reads
\begin{equation}
    M_2\times Y \rightarrow M_3\times \partial Y
\end{equation}
where $M_2$ is the 2D spacetime where the D1-branes worldvolume extending, living on the boundary of a 3D bulk manifold $M_3= M_2\times \mathbb{R}_{\geq 0}$. The asymptotic boundary $\partial Y$ is given by the Sasaki-Einstein base space of $Y$.

To compute a SymTFT supported on $M_3$, we ought to dimensionally reduce the topological sector of IIB string theory on $\partial Y$. The relevant topological action involves various IIB fluxes, among which we are interested in $F_5$, $G_3$, and $H_3$ which are electrically (resp. magnetically) sourced by D3 (resp. D3), D1 (resp. D5), and F1 (resp. NS5)-branes, respectively\footnote{In this work we consider the IIB string theory background without 7-branes.}. Technically, the dimensional reduction starts from treating the various IIB supergravity fluxes as elements
in differential cohomology classes $\breve{H}^*(\partial Y)$ (see, e.g., \cite{Freed:2006ya, Freed:2006yc, Belov:2006jd}) sitting in the exact sequences
\begin{equation}
	0~\rightarrow ~H^{p-1}(\partial Y;\R/\Z)~\rightarrow ~\breve{H}^{p}(\partial Y)~\rightarrow ~\Omega_\Z^p(\partial Y)~\rightarrow ~0
\end{equation}
and
\begin{equation}\label{eq: short exact sequence for diff coho}
    0~\rightarrow ~\Omega^{p-1}(\partial Y)/\Omega^{p-1}_{\Z}(\partial Y)~\rightarrow ~\breve{H}^p(\partial Y) ~\stackrel{\pi}{\rightarrow} ~H^{p}(\partial Y; \Z) ~\rightarrow ~0
\end{equation}
where $\Omega^{p-1}(\partial Y)$ (resp. $\Omega^{p-1}_{\Z}(\partial Y)$) denotes closed differential $p$-forms (resp. with integral periods).\footnote{The first sequence is physically referred to as the field-strength sequence, while the second one is for the characteristic class. We refer the reader to \cite{Freed:2006yc} for more details.} To be explicit, we assume the following cohomology classes for $\partial Y$:
\begin{equation}\label{eq: coho classes for 7D SE}
    H^*(\partial Y;\mathbb{Z})=\left\{\mathbb{Z},0,\mathbb{Z}^{b_2}\oplus \Gamma_2,0, \Gamma_4, \mathbb{Z}^{b_5},\Gamma_{6},\mathbb{Z} \right\},
\end{equation}
which covers an infinite family of Calabi-Yau 4-fold singularities, including orbifolds $\C^4/\Gamma$ with isolated singularities, as well as cones over $Y^{p,k}(B_4)$ with base space $B_4=\mathbb{P}^1\times \mathbb{P}^1$ or $B_4=\mathbb{P}^2$ \cite{Martelli:2008rt}. In the above cohomology classes, $b_2$ and $b_5$ are, respectively, the second and the fifth Betti numbers, and $\Gamma_n$ is the torsional part of the $n$-th cohomology class. In particular, we have $b_2=b_5$ and $\Gamma_6=\Gamma_2$ due to Poincar\'{e} duality and universal coefficient theorem \cite{hatcher2002algebraic}.\footnote{In principle, one can perform the analysis for any internal geometries with conical singularities, even for non-Calabi-Yau geometries. For example, it would be interesting to investigate the symmetry structure for exotic 2D (0,1) QFTs associated to Spin(7) backgrounds \cite{Franco:2021ixh, Franco:2021vxq}, which we defer to future work.}

Due to the self-duality of the 5-form field strength $F_5$, a consistent way to capture the relevant terms in IIB supergravity action is to write down a topological action defined on an auxiliary 11D manifold $M_{11}$ (see, e.g., \cite{Bah:2020uev, Hsieh:2020jpj, GarciaEtxebarria:2024fuk, Belov:2006jd, Belov:2006xj, Monnier:2012xd, Heckman:2017uxe, Apruzzi:2023uma, Lawrie:2023tdz, Yu:2023nyn})
\begin{equation}\label{eq: 11D top action}
	\frac{S_{11}}{2\pi}= \int_{M_{11}}\frac{1}{2}F_5\wedge dF_5-F_5\wedge H_3\wedge G_3
\end{equation}
and then perform the cohomology uplift \cite{Freed:2006yc, Lawrie:2023tdz, Yu:2023nyn, GarciaEtxebarria:2024fuk}
\begin{equation}\label{eq: 11D diff coho action}
    \frac{S_{11}}{2\pi} \rightarrow \int_{M_{11}}\frac{1}{2}\breve{F}_6\star \breve{F}_6-\breve{F}_6\star \breve{H_3} \star \breve{G}_3.
\end{equation}
In the above expression, IIB fluxes are uplifted into differential cohomology elements\footnote{There are some subtitles with this auxiliary 11D formalism under differential cohomology uplift. See \cite{GarciaEtxebarria:2024fuk}}
\begin{equation}
    \breve{F}_6\in \breve{H}^6(M_{11}), \breve{H}_3\in \breve{H}^3(M_{11}), \breve{G}_3\in \breve{H}^3(M_{11}),
\end{equation}
and the $\star$ symbol defines a bilinear product operation on Cheeger--Simons characters  $\breve{H}^{k_1}(M_d)\times \breve{H}^{k_2}(M_d)=\breve{H}^{k_1+k_2}(M_d)$ \cite{brylinski2007loop, Cheeger1985DifferentialCA}. In particular, the integral (\ref{eq: 11D diff coho action}) describes a perfect pairing $\breve{H}^{k_1}(M_d)\times \breve{H}^{d+1-k_1}(M_d)\rightarrow \mathbb{R}/\mathbb{Z}$.\footnote{We refer the reader to \cite{Freed:2006ya, Freed:2006yc, Hsieh:2020jpj, Apruzzi:2021nmk} for physical perspective review of differential cohomology and \cite{baer2013differential} for a more mathematical one.} The case of our interest is $M_{11}=N_4\times \partial Y$, where the auxiliary 4-manifold $N_4$ has the boundary $\partial N_4=M_3$ supporting the SymTFT to be derived. Under this decomposition of $M_{11}$, we can expand the $\breve{F}_6, \breve{H}_3$ and $\breve{G}_3$ based on (\ref{eq: coho classes for 7D SE})
\begin{equation}\label{eq: expansion of diff cochains}
\begin{split}
     \breve{F}_{6}=&\sum_{j}\breve{a}_4^{(j)}\star \breve{t}_{2(j)}+\sum_{i}\breve{a}_2^{(i)}\star \breve{t}_{4(i)}+\cdots,\\
     \breve{H}_{3}=&\sum_j\breve{b}_{1}^{(j)}\star \breve{t}_{2(j)}+\cdots,\\
    \breve{G}_{3}=&\sum_j\breve{c}_1^{(j)}\star \breve{t}_{2(j)}+\cdots,
\end{split}
\end{equation}
where $\breve{t}_{2(j)}$ and $\breve{t}_{4(i)}$ are generators respectively corresponding to $\Gamma_2$ and $\Gamma_4$, with indices $j$ and $i$ run over the generators. In the above expansion (\ref{eq: expansion of diff cochains}), we only make the torsional part explicit while suppressing the non-torsional part in ``$\cdots$''. This is because we only focus on finite symmetries in this work. We defer the investigation of the non-torsional part and its associated continuous symmetries to future work.

Performing the dimensional reduction via integrating over $\partial Y$, we arrive at a 4D action 
\begin{equation}\label{eq: 3D TFT part 1}
\begin{split}
    \frac{S_{4}}{2\pi}&=\int_{N_4}\frac{1}{2}\sum_{i_1,i_2}\Lambda_{i_1i_2}\breve{a}_2^{(i_1)}\star \breve{a}_2^{(i_2)}+ \sum_{i,j_1,j_2}\Delta_{ij_1j_2} \breve{a}_2^{(i)}\star \breve{b}_1^{(j_1)} \star \breve{c}_1^{(j_2)},\\
\end{split}
\end{equation}
where $\Lambda_{i_1i_2}$ and $\Delta_{ij_2j_2}$ are linking numbers defined by
\begin{equation}\label{eq: linking numbers}
\begin{split}
     \Lambda_{i_1i_2}&\equiv \int_{\partial Y}\breve{t}_{4(i_1)}\star \breve{t}_{4(i_2)}~\text{mod 1},\\
 \Delta_{ij_1j_2}&\equiv -\int_{\partial Y}\breve{t}_{4(i)}\star \breve{t}_{2(j_1)} \star \breve{t}_{2(j_2)}~\text{mod 1}.
\end{split}
\end{equation}
These linking numbers can be derived from intersection numbers between divisors of the Calabi-Yau 4-fold $Y$ \cite{vanBeest:2022fss, Yu:2023nyn}.\footnote{For toric Calabi-Yau manifolds, intersection numbers between various divisors can be computed following, e.g., Chapter 7 of \cite{hori2003mirror}.} We further assume $M_3=\partial N_4$ satisfying $H^2(M_3,\mathbb{Z})=0$ so that $\breve{a}_2$ supported on $N_4$ can be trivialized to $\breve{a}_1$ on $M_3$ \cite{GarciaEtxebarria:2024fuk}. This allows us to write down a 3D action in terms of ordinary cochains
 \begin{equation}\label{eq: middle step of 3D SymTFT}
     \frac{S_3}{2\pi}=\int_{M_3}\frac{1}{2}\Lambda_{i_1i_2}a_1^{(i_1)}\cup \delta a_1^{(i_2)}+\sum_{i,j_1,j_2}\Delta_{ij_1j_2} a_1^{(i)}\cup b_1^{(j_1)} \cup c_1^{(j_2)}.
 \end{equation}

It is easy to see from (\ref{eq: middle step of 3D SymTFT}) that there are three classes of 0-form finite symmetries $\Gamma^{(a^{(i)})}, \Gamma^{(b^{(j)})}$ and $\Gamma^{(c^{(j)})}$, with background fields $a^{(i)}, b^{(j)}$ and $c^{(j)}$, respectively. The geometric counterparts of these three symmetries are
\begin{equation}\label{eq: defect group part 1}
	\Gamma^{(a^{(i)})}=\Gamma_4^{(i)}, ~\Gamma^{(b^{(j)})}=\Gamma_2^{(j)}, ~\Gamma^{(c^{(j)})}=\Gamma_2^{(j)},
\end{equation}
where $\Gamma_2=\oplus_i\Gamma^{(i)}_2$ and $\Gamma_4=\oplus_j\Gamma^{(j)}_4$ are torsional cohomology generators of $H^*(\partial Y; \Z)$ in (\ref{eq: coho classes for 7D SE}). However, the action (\ref{eq: middle step of 3D SymTFT}) is incomplete. This can be seen by noticing that for $\Gamma^{(a^{(i)})}$ symmetries, the first term in (\ref{eq: middle step of 3D SymTFT}) encodes the dual quantum symmetry information (if possible gauging is allowed) since it is single-derivative in the form of Dijkgraaf-Witten/Chern-Simons. Similarly, we need the information about the possible gauging of $\Gamma^{(b^{(j)})}$ and $\Gamma^{(c^{(j)})}$ and the dual quantum symmetries, for which the single-derivative terms are missing in (\ref{eq: middle step of 3D SymTFT}).

According to \cite{Yu:2023nyn} (see also, e.g., \cite{Apruzzi:2021nmk, Baume:2023kkf} for a similar discussion in higher-dimensional setups), these extra terms for $\Gamma^{(b^{(j)})}$ and $\Gamma^{(c^{(j)})}$ symmetries can be added by the following argument. Notice that the first term in (\ref{eq: middle step of 3D SymTFT}) is single-derivative and comes from the linking between the following two wrapped D3-branes 
\begin{equation}\label{eq: D3-D3 pairings}
	\text{D3-brane on $\gamma_3^{(i_1)}$, D3-brane on $\gamma_3^{(i_2)}$},
\end{equation}
where $\gamma_3^{(i_1)}$ and $\gamma_3^{(i_2)}$ are torsional 3-cycles dual to two generators of the 
cohomology class $\text{Tor}{H}^{4}(\partial Y;\Z)=\Gamma_4$. The corresponding differential cohomology generators are $\breve{t}_{4(i_1)}$ and $\breve{t}_{4(i_1)}$. The linking invariant between these two torsional 3-cycles is encoded in the first linking pairing in (\ref{eq: linking numbers}). Similarly, one can consider the linking pairs between the following wrapped branes
\begin{equation}\label{eq: F1-NS5 and D1-D5 pairings}
\begin{split}
	&\text{F1-string on $\gamma_1^{(j_1)}$, NS5-brane on $\gamma_{5(j_2)}$},\\
	&\text{D1-string on $\gamma_1^{(j_1)}$, D5-brane on $\gamma_{5(j_2)}$},
\end{split}
\end{equation}
where $\gamma_1^{(j_1)}$ and $\gamma_{5(j_2)}$ are torsional 1-cycle and 5-cycle dual to the cohomology class $\text{Tor}H^2(\partial Y, \Z)=\Gamma_2$ and $\text{Tor}H^6(\partial Y, \Z)=\Gamma_6$, respectively. The linking invariant between these two torsional cycles can be computed via the differential cohomology pairing
\begin{equation}\label{eq: linking number 2}
	\Omega_{j_1}^{j_2}\equiv \int_{\partial Y}\breve{t}_{2(j_1)}\star \breve{t}_{6}^{(j_2)}~\text{mod}~1 ,
\end{equation}
where $\breve{t}_{6}^{(j_2)}$ is the $j_2$-th generator of $\breve{H}^6(\partial Y)$ associated to $\Gamma_6=\oplus_{j}\Gamma_{6(j)}$. These give rise to two more classes of 0-form finite symmetries 
\begin{equation}\label{eq: defect group part 2}
	\Gamma^{(\hat{b}_{(j)})}=\Gamma_{6(j)}, ~ \Gamma^{(\hat{c}_{(j)})}=\Gamma_{6(j)}
\end{equation}
whose background fields are $\hat{b}_{(j)}$ and $\hat{c}_{(j)}$, corresponding to correspond to NS5 and D5-branes in (\ref{eq: F1-NS5 and D1-D5 pairings}) respectively, in obvious notations. 

Adding the extra terms from the pairing (\ref{eq: F1-NS5 and D1-D5 pairings}) and (\ref{eq: linking number 2}), we end up with the 3D SymTFT action
\begin{equation}\label{eq: SymTFT for 4-fold}
\boxed{
\begin{split}
	S_3=&2\pi \int_{M_3}\frac{1}{2}\Lambda_{i_1i_2}a_1^{(i_1)}\cup \delta a_1^{(i_2)}+\Omega_{j_1}^{j_2}b_1^{(j_1)}\cup \hat{b}_{1(j_2)}-\Omega_{j_1}^{j_2}c_1^{(j_1)}\cup \hat{c}_{1(j_2)}+\\
	&+\sum_{i,j_1,j_2}\Delta_{ij_1j_2} a_1^{(i)}\cup b_1^{(j_1)} \cup c_1^{(j_2)}.
\end{split}
	}
\end{equation}

Before investigating this general SymTFT, we remark that though in this work, we derive the single-derivative terms in (\ref{eq: SymTFT for 4-fold}) from the flux non-commutativity and the associated brane linking in (\ref{eq: D3-D3 pairings}) and (\ref{eq: F1-NS5 and D1-D5 pairings}), there is an alternative way to derive these terms via dimensionally reducing the kinetic terms directly in 10D IIB supergravity action in terms of the non-harmonic differential forms (see e.g., \cite{Baume:2023kkf,Camara:2011jg}).

\subsubsection{Defect Group and Relativeness}

Let us start with the defect group. The defect group for the SymTFT (\ref{eq: SymTFT for 4-fold}) can be read from (\ref{eq: defect group part 1}) and (\ref{eq: defect group part 2}) 
\begin{equation}\label{eq: defect group for the full TFT}
\begin{split}
	\mathbb{D}&=\prod_i\Gamma^{(a^{(i)})}\times \prod_j\left(\Gamma^{(b^{(j)})}\times \Gamma^{(\hat{b}_{(j)})}\right)\times \prod_j \left(\Gamma^{(c^{(j)})}\times \Gamma^{(\hat{c}_{(j)})}\right)\\
	&=\Gamma_4\times (\Gamma_2\times \Gamma_6)\times (\Gamma_2\times \Gamma_6),
\end{split}
\end{equation}
where the $\Gamma$'s are given by the geometric data (\ref{eq: coho classes for 7D SE}). Using the condition $\Gamma_2=\Gamma_6$ below (\ref{eq: coho classes for 7D SE}), we obtain the defect group
\begin{equation}
	\mathbb{D}=\Gamma_4\times (\Gamma_2^{(b)}\times \Gamma_2^{(\hat{b})})\times (\Gamma_2^{(c)}\times \Gamma_2^{(\hat{c})}),
\end{equation}
where we use indices to distinguish various $\Gamma_2$ factors.

The Dirac pairing for the defect group, as we discussed around (\ref{eq: symtft for dirac pairing}), is given by the single-derivative terms in the SymTFT, i.e., the first line in (\ref{eq: SymTFT for 4-fold}). It is easy to see that for the $\Gamma_2$ part of the defect group, there always exist Lagrangian subgroups in the form of 
\begin{equation}\label{eq: lagrangian subgroup of bc part}
	\Gamma_2\times \Gamma_2\subset (\Gamma_2^{(b)}\times \Gamma_2^{(\hat{b})})\times (\Gamma_2^{(c)}\times \Gamma_2^{(\hat{c})}),
\end{equation}
associated with purely ``electric'' or ``magnetic'' gapped boundary conditions. For example, the subgroup $\Gamma_2^{(\hat{b})}\times \Gamma_2^{(\hat{c})}$ corresponds to the gapped boundary condition
\begin{equation}\label{eq: D boundary condition for b and c}
	b^{(j)}~\text{and}~c^{(j)}~\text{Dirichlet}, ~\hat{b}_{(j)}~\text{and}~\hat{c}_{(j)}~\text{Neumann}
\end{equation}
for all $j$. This matches the fact that for this part of the defect group, the Dirac pairing is anti-symmetric, analogous to the $\mathfrak{su}(2)$ SYM example in Section 2.1.

However, the $\Gamma_4$ part in the defect group is not guaranteed to have a Lagrangian subgroup. Recall one necessary condition for a Lagrangian subgroup is (\ref{eq: necessary condition for L}), which means only if the $|\Gamma_4|=m^2, m\in \Z$ can a Lagrangian subgroup of $\Gamma_4$ possibly exist. In other words, if $|\Gamma_4|$ is not a complete square, then there is no well-defined gapped boundary condition that can be picked for $a^{(i)}$ fields. Intuitively, this can be understood since the single-derivative terms for $a^{(i)}$ in the SymTFT (\ref{eq: SymTFT for 4-fold}) may just include Chern-Simons-type terms similar to (\ref{eq: 4l+3 CS theory}) where no gapped boundary condition can be defined. According to the discussion in Section 2.2, in these cases, the associated 2D QFTs are intrinsically relative. Therefore, we end up with the following statement:

\paragraph{A sufficient condition for relativeness:} \emph{Let $\mathcal{T}$ be a 2D QFT engineered from a Calabi-Yau 4-fold $Y$ probed by D1-branes. Then $\mathcal{T}$ is an intrinsically relative QFT if $|\text{Tor}H^4(\partial Y;\Z)|$ is not a complete square.}\label{sta: 1}

\vspace{0.3cm}
This relativeness for 2D QFTs is reminiscent of an infinite class of 6D SCFTs as relative theories \cite{DelZotto:2015isa, GarciaEtxebarria:2019caf, Gukov:2020btk, Lawrie:2023tdz}. In fact, the relativeness of 6D SCFTs and our interested 2D QFTs enjoy a related string theory origin, namely the flux-noncommutativity for the self-dual $F_5$ flux in IIB string theory. After dimensional reduction to the topological sector of IIB compactification, this flux-noncommutativity leads to the obstruction of a well-defined boundary condition ``at infinity" for lower-dimensional fields inherited from $F_5$. In the SymTFT language, this translates into the self-linking of the operators built on D3-branes and, as a result, prevents any Lagrangian subalgebra/gapped boundary condition. We will illustrate this top-down approach to the 2D relativeness with an explicit example in Section 5.
\subsubsection{Non-invertible Symmetries and their Anomalies}
After discussing the case when 2D QFTs are intrinsically relative, let us now consider the case when Calabi-Yau 4-folds $Y$ with the order of $\Gamma_4=\text{Tor}H^4(\partial Y; \Z)$ a complete square, and further assume the SymTFT (\ref{eq: SymTFT for 4-fold}) admits gapped boundary conditions. We can split the $\Gamma_4$ generators as 
\begin{equation}\label{}
    \Gamma_4=\oplus_i\Gamma_4^{(i)}=\oplus_k\Gamma_4^{(k)}\oplus_l \Gamma_4^{(l)},
\end{equation}
and then, without loss of generality, assume a standard ``electric'' gapped boundary condition, picking a Dirichlet condition for the following fields
\begin{equation}\label{eq: generic D condition for 3D SymTFT}
    a_1^{(k)}, b_1^{(j)}, c_1^{(j)}~\text{Dirichlet}, 
\end{equation}
The resulting absolute QFT enjoys three classes of invertible finite symmetries
\begin{equation}\label{eq: electric polarization symmetry}
G^{(a^{(k)})}=\Gamma_4^{(k)},~G^{(b^{(j)})}=\Gamma_2^{(j)},~G^{(c^{(j)})}=\Gamma_2^{(j)},
\end{equation}
whose respective background gauge fields are those given in (\ref{eq: generic D condition for 3D SymTFT}). 

For these three classes of invertible symmetries, there are mixed anomalies from the relevant cubic terms in the SymTFT (\ref{eq: SymTFT for 4-fold})
\begin{equation}
    \int_{M_3}\sum_{k,j_1,j_2}\Delta_{kj_1j_2} a_1^{(k)}\cup b_1^{(j_1)} \cup c_1^{(j_2)}.
\end{equation}
According to \cite{Kaidi:2021xfk}, which we also reviewed in Section 2.3, gauging two of the three classes of symmetries will promote the leftover one to non-invertible symmetries. In fact, based on our discussion around (\ref{eq: lagrangian subgroup of bc part}), it is always possible to gauge the $G^{(b^{(j)})}$ and $G^{(c^{(j)})}$ symmetry and end up with a ``magnetic'' boundary condition 
\begin{equation}\label{eq: generic magnetic D condition for 3D SymTFT}
    a_1^{(k)}, \hat{b}_{1(j)}, \hat{c}_{1(j)}~\text{Dirichlet}.
\end{equation}
The resulting absolute QFT under this condition has two classes of invertible symmetries
\begin{equation}\label{eq: quantum symmetries for b and c}
    G^{(\hat{b}_{(j)})}=\Gamma_2^{(j)}, G^{(\hat{c}_{(j)})}=\Gamma_2^{(j)}
\end{equation}
as quantum symmetries from gauging $G^{(b^{(j)})}$ and $G^{(c^{(j)})}$. Furthermore, it also has non-invertible symmetries whose topological lines obey the fusion rule
\begin{equation}\label{eq: fusion rule of general non-invertible symmetries}
\boxed{
\begin{split}
    &\mathcal{N}_{(k)}\otimes \mathcal{N}_{(k)}=\oplus_g \eta^{(g)}_{(b^{(j)})}\oplus_g \eta^{(g)}_{(c^{(j)})},\\
    &\mathcal{N}_{(k)}\otimes \eta^{(g)}_{(b^{(j)})}=\eta^{(g)}_{(b^{(j)})}\otimes \mathcal{N}_{(k)}= \mathcal{N}_{(k)},\\
    &\mathcal{N}_{(k)}\otimes \eta^{(g)}_{(c^{(j)})}=\eta^{(g)}_{(c^{(j)})}\otimes \mathcal{N}_{(k)}= \mathcal{N}_{(k)},\\
    &\eta^{(g)}_{(b^{(j)})}\otimes \eta^{(h)}_{(b^{(j)})}=\eta^{(gh)}_{(b^{(j)})},\\
    &\eta^{(g)}_{(c^{(j)})}\otimes \eta^{(h)}_{(c^{(j)})}=\eta^{(gh)}_{(c^{(j)})},
\end{split}
    }
\end{equation}
where $\eta^{(g)}_{b^{(j)}}$ (resp. $\eta^{(g)}_{c^{(j)}}$) is the symmetry operator corresponding to the group element $g$ of $G^{(b^{(j)})}$ (resp. $\eta^{(g)}_{c^{(j)}}$). The right-hand side of the first line in the above fusion rule is the sum over all symmetry operators for $G^{(b^{(j)})}$ and $G^{(c^{(j)})}$, which is the condensation defect \cite{Roumpedakis:2022aik} for 1-gauging of the $G^{(b^{(j)})}\times G^{(c^{(j)})}$ symmetry. Using the identification of the $G^{(b^{(j)})}\times G^{(c^{(j)})}$ to the geometric torsional group $\Gamma_2^{(j)}$, we realize the categorical symmetry generated by the fusion rules (\ref{eq: fusion rule of general non-invertible symmetries}) is the TY$(\Gamma_2^{(j)}\times \Gamma_2^{(j)})$ fusion category \cite{TAMBARA1998692}.

\subsubsection*{Anomalies}

It is then natural to ask whether these non-invertible symmetries are anomalous. The relevant single-derivative terms in the SymTFT (\ref{eq: SymTFT for 4-fold}) read 
\begin{equation}
    2\pi \int_{M_3}\frac{1}{2}\Lambda_{k_1k_2}a_1^{(k_1)}\cup \delta a_1^{(k_2)}.
\end{equation}
It is easy to see that when the diagonal element of $\Lambda_{k_1k_2}$ is non-vanishing (mod 1), the invertible $G^{(a^{(k)})}$ symmetry has a self-anomaly. Under the Dirichlet condition (\ref{eq: generic D condition for 3D SymTFT}) for $a_1^{(k)}$, this self-anomaly can be expressed in terms of the background field profile $A_1^{(k)}$ as
\begin{equation}
    2\pi \int_{M_3}\frac{1}{2}\Lambda_{kk}A_1^{(k)}\cup \delta A_1^{(k)}.
\end{equation}
For example, the self-anomaly in (\ref{eq: anomaly inflow}) for $\Z_2$ symmetry corresponds to the  case when $A_1^{(k)}$ is a $\Z_2$ background gauge field and $\Lambda_{kk}=\frac{1}{4}$.

The non-invertible symmetry generated by $\mathcal{N}_{(k)}$ in (\ref{eq: fusion rule of general non-invertible symmetries}) is derived from promoting the invertible $G^{a^{(k)}}$ symmetry. According to \cite{Kaidi:2023maf} and our discussion in Section 2.3, these non-invertible symmetries will also inherit the self-anomalies of $G^{(a^{(k)})}$. From the topological operator perspective, the invertible symmetry $G^{a^{(k)}}$ and its non-invertible symmetry promotion are generated by a certain topological operator $L_{(k)}$ in the SymTFT bulk, while the invertibility (resp. non-invertibility) of the symmetry is determined by the behavior of the operator under the gapped boundary conditions (\ref{eq: generic D condition for 3D SymTFT}) (resp. (\ref{eq: generic magnetic D condition for 3D SymTFT})):
\begin{equation}\label{eq: line behavior under different conditions}
    L_{(k)}\rightarrow
\left\{ 
\begin{array}{cc}
     D_{(k)}, & \text{under (\ref{eq: generic D condition for 3D SymTFT})} \\
     \mathcal{N}_{(k)}, & \text{under (\ref{eq: generic magnetic D condition for 3D SymTFT})}
\end{array}
\right.
\end{equation}
where $D_{(k)}$ is the invertible symmetry operator for $G^{(a^{(k)})}$. The self-anomaly for the $G^{(a^{(k)})}$ symmetry and the non-invertible symmetry generated by $\mathcal{N}_{(k)}$ then shares the same origin, namely the topological line operator $L_{(k)}$ in the SymTFT is not endable on the gapped boundary, implying the obstruction to the gauging. We will illustrate this idea via explicit examples in Section 5. 

We conclude this subsection by the following remarks. The derivation for the SymTFTs and categorical symmetries we have done so far is purely via the geometric data. A natural question is whether the same result can be derived from the 2D field-theory information via brane brick models. In a similar stringy setup, namely 4D QFTs on D3-branes probing Calabi-Yau 3-folds, it was recently shown in \cite{Braeger:2024jcj} that the (co)homology data for finite global symmetries can be directly derived from the quiver data. We believe the (co)homology data in (\ref{eq: coho classes for 7D SE}) can be similarly computed from the quivers and J- and E-terms. Indeed, it is computed in, e.g., \cite{Franco:2015tya} that the generators of Calabi-Yau 4-folds can be derived explicitly in terms of the gauge-invariant chiral operators of 2D QFTs. It would be interesting to further study how to derive the geometric data determining the categorical symmetries, including the Betti numbers $b_{2}$ and $b_5$, from the field-theory information, which we leave this for future work.

\subsection{Topological Operators from Branes}

As we reviewed in Section 2, starting with a (D+1)-dimensional SymTFT, the relativeness and the categorical symmetry structure for the associated D-dimensional QFT can be investigated via the topological operators and their linkings in the SymTFT bulk. In principle, after obtaining the general SymTFT (\ref{eq: SymTFT for 4-fold}), which we also rewrite here for the ease of reading
\begin{equation}
\boxed{
\begin{split}
	S_3=&2\pi \int_{M_3}\frac{1}{2}\Lambda_{i_1i_2}a_1^{(i_1)}\cup \delta a_1^{(i_2)}+\Omega_{j_1}^{j_2}b_1^{(j_1)}\cup \hat{b}_{1(j_2)}-\Omega_{j_1}^{j_2}c_1^{(j_1)}\cup \hat{c}_{1(j_2)}+\\
	&+\sum_{i,j_1,j_2}\Delta_{ij_1j_2} a_1^{(i)}\cup b_1^{(j_1)} \cup c_1^{(j_2)} ,
\end{split}
	}
\end{equation}
one can try to perform the analysis of deriving topological operators from a purely field-theoretic perspective. However, given that SymTFT enjoys a string theory embedding, it is natural to expect there are also stringy counterparts for their topological operators. We will show in this subsection that this is indeed the case. Following the idea introduced in \cite{Heckman:2022muc, Apruzzi:2022rei, GarciaEtxebarria:2022vzq, Heckman:2022xgu} (see also, e.g., \cite{Apruzzi:2023uma, Yu:2023nyn, Etheredge:2023ler, Bah:2023ymy, Cvetic:2023plv, Baume:2023kkf}), we will present a top-down approach to topological defect line operators within the above SymTFTs, where the 1D TFT action living on the line operator is derived from the topological sector of the brane worldvolume action via dimensional reduction. 

Let us consider D3-branes, coupled to the self-dual $F_5$ flux, which is the IIB origin of the $a^{(i)}$ fields in the SymTFT. The topological sector of a D3-brane worldvolume theory is given by the Wess-Zumino term \cite{Douglas:1995bn}
\begin{equation}\label{eq: WZ for D3}
    S_{\text{D3}}^{\text{WZ}}=\int \D v_1 \exp\left( 2\pi i \int_{M_4}C_4+C_2\wedge (dv_1-B_2)+\cdots \right)
\end{equation}
where we only write explicitly terms relevant to our following discussion and suppress other terms into $\cdots$. In the above expression, $v_1$ is the dynamical $U(1)$ gauge field from the open string fluctuation, while $C_4$, $C_2$ and $B_2$ are closed string sector background fields, electrically coupled to D3-brane, D1-, and F1-string charges. 

Due to the topological nature of the Wess-Zumino terms, it is certain that the dimensional reduction of (\ref{eq: WZ for D3}) will give rise to a topological operator. However, Wess-Zumino terms do not capture all the topological effects of the dimensionally reduced D3-brane physics. Recall that in the derivation of the SymTFT (\ref{eq: SymTFT for 4-fold}), the cubic terms come from the topological sector $\int C_4\wedge dB_2\wedge dC_2$ of the IIB string theory, but the single-derivative terms are inherited from the kinetic terms, which are \emph{not} topological in 10D. Similarly, in the case of reducing the brane worldvolume action, there are also lower-dimensional topological effects captured by the Dirac-Born-Infeld action \cite{Leigh:1989jq}. The relevant part for us is the kinetic term for the dynamical $U(1)$ field from the open string sector,
\begin{equation}
    \int_{M_4}dv_1\wedge \star_{M_4} (dv_1).
\end{equation} Including the effect of this kinetic term, one can write down a generalized topological action defined on an auxiliary 5-manifold $N_5$ with $\partial N_5=M_4$ (see, e.g., \cite{Apruzzi:2023uma, Yu:2023nyn})
\begin{equation}
    S_{\text{D3}}^{\text{top}}=\int \D \hat{f}_2 \D f_2 \exp\left( 2\pi i \int_{N_5}F_5+\hat{f}_2\wedge df_2 +G_3\wedge (f_2-B_2)\right).
\end{equation}
Note that now $f_2, F_5$ and $G_3$ are all $U(1)$ \emph{connections} in $N_5$ instead of field-strengths. 

Now consider the case we are interested in, namely D3-brane wrapping on a torsional 3-cycle $\gamma_3^{(i)}$, corresponding to the cohomology $H^4(\partial Y; \Z)$. Similarly to the dimensional reduction of the IIB supergravity discussed in Section 4.1, we uplift the brane action in terms of differential cohomology  
\begin{equation}\label{eq: diff coho action for D3}
    S_{\text{D3}}^{\text{top}}\rightarrow \int \D \breve{\hat{f}}_3 \D \breve{f}_3 \exp\left( 2\pi i \int_{N_2\times \gamma_3^{(i)}}\breve{F}_6+\breve{\hat{f}}_3\star \breve{f}_3 +\breve{G}_3\star (\breve{f}_3-\breve{H}_3)\right).
\end{equation}
Expanding $\breve{\hat{f}}_3$ and $\breve{f}_3$ as 
\begin{equation}
\begin{split}
    &\breve{\hat{f}}_3=\sum_j \breve{\hat{\phi}}_1^{(j)}\star \breve{t}_{2(j)}+\cdots,\\
    &\breve{f}_3=\sum_j\breve{\phi}_1^{(j)}\star \breve{t}_{2(j)}+\cdots,
\end{split}
\end{equation}
together with the expansion for $\breve{F}_6, \breve{G}_3$ and $\breve{H}_3$ given in (\ref{eq: expansion of diff cochains}), we can dimensionally reduce (\ref{eq: diff coho action for D3}) into 
\begin{equation}
    \int \prod_j \D \breve{\hat{\phi}}_1^{(j)} \D \breve{\phi}_1^{(j)}\exp \left( 2\pi i \int_{N_2}\sum_{i'}\Lambda_{ii'}\breve{a}_2^{(i')}-
    \sum_{j_1,j_2}\Delta_{ij_1j_2} \left( \breve{\hat{\phi}}_1^{(j_1)}\star \breve{\phi}_1^{(j_2)}+\breve{c}_1^{(j_1)}\star (\breve{\phi}_1^{(j_2)}-\breve{b}_1^{(j_2)})\right) \right).
\end{equation}
Using $\partial N_2=M_1$, we reduce the result in terms of ordinary cochains and write down the following topological line operator 
\begin{equation}\label{eq: general line operator from D3}
\boxed{
\begin{split}
    L_{(i)}^{\text{D3}}&=\exp\left(2\pi i \int_{M_1} \sum_{i'}\Lambda_{ii'}a_1^{(i')} \right)\\
    &\times \int \prod_j \D \hat{\phi}_0^{(j)} \D \phi_0^{(j)}\exp \left(- 2\pi i \int_{M_1} \sum_{j_1,j_2}\Delta_{ij_1j_2}\left(\hat{\phi}_0^{(j_1)}\cup \delta \phi_{0}^{(j_2)}+c_1^{(j_1)}\cup \phi_0^{(j_2)}-b_1^{(j_2)}\cup \hat{\phi}_0^{(j_1)}  \right) \right)
\end{split}
}
\end{equation}
from D3-brane on $\gamma_3^{(i)}$, where $\Lambda_{ii'}$ and $\Delta_{ij_1j_2}$ are linking numbers defined in (\ref{eq: linking numbers}). Recall that $\phi_0^{(j)}$ and $\hat{\phi}_0^{(j)}$ are from the expansion via $\oplus_j\Gamma_2^{(j)}=\Gamma_2=\text{Tor}H^2(\partial Y; \Z)$ generators, it is easy to see the above line operator is a stacking of an invertible line operator (the first line in (\ref{eq: general line operator from D3})) and a non-invertible line built from a path integral over 1D twisted $\Gamma_2^{(j_1)}\times \Gamma_2^{(j_2)}$ Dijkgraaf-Witten theories (the second line in (\ref{eq: general line operator from D3})).

Similarly, one can derive line operators from respectively a F1-string and a D1-string wrapping on torsional 1-cycle $\gamma_1^{(j)}$, for which the result simply is 
\begin{equation}\label{eq: general line operators for F1 and D1}
\boxed{
    L_{(j)}^{\text{F1}}=\exp \left( 2\pi i \int_{M_1}\sum_{j'}\Omega_{j'}^jb_1^{j'} \right),~ L_{(j)}^{\text{D1}}=\exp \left( 2\pi i \int_{M_1}\sum_{j'}\Omega_{j'}^jc_1^{j'} \right).
    }
\end{equation}
which are invertible line operators.

\subsubsection{Revisiting Relativeness, Non-invertible Symmetries and their Anomalies}

The relativeness, non-invertible symmetries, and their anomalies discussed in Section 4.1 can now be revisited from the topological operators' perspective. Here, we briefly present a rough discussion and leave the explicit investigation with examples in Section 5. 

The possible intrinsic relativeness for the 2D QFT is due to the linking 
\begin{equation}\label{eq: linking a lines}
    \langle L_{(i_1)}^{\text{D3}}(M_1) L_{(i_2)}^{\text{D3}}(M_1') \rangle,
\end{equation}
which in general is non-trivial due to the SymTFT term $2\pi \int_{M_3}\frac{1}{2}\Lambda_{i_1i_2}a_1^{(i_1)}\cup \delta a_1^{(i_2)}$, similarly to (\ref{eq: linking of non-invertible lines in KNZZ}) (see also Appendix B in \cite{Kaidi:2023maf}). This non-trivial linking can prevent the existence of a maximally isotropic subspace for $\{L_{(i)}^{\text{D3}}\}$, which aligns with the case when $\Gamma_4$ does not have a Lagrangian subgroup, as discussed in Section 4.1.1.

\begin{figure}[h]
    \centering
    \includegraphics[width=14cm]{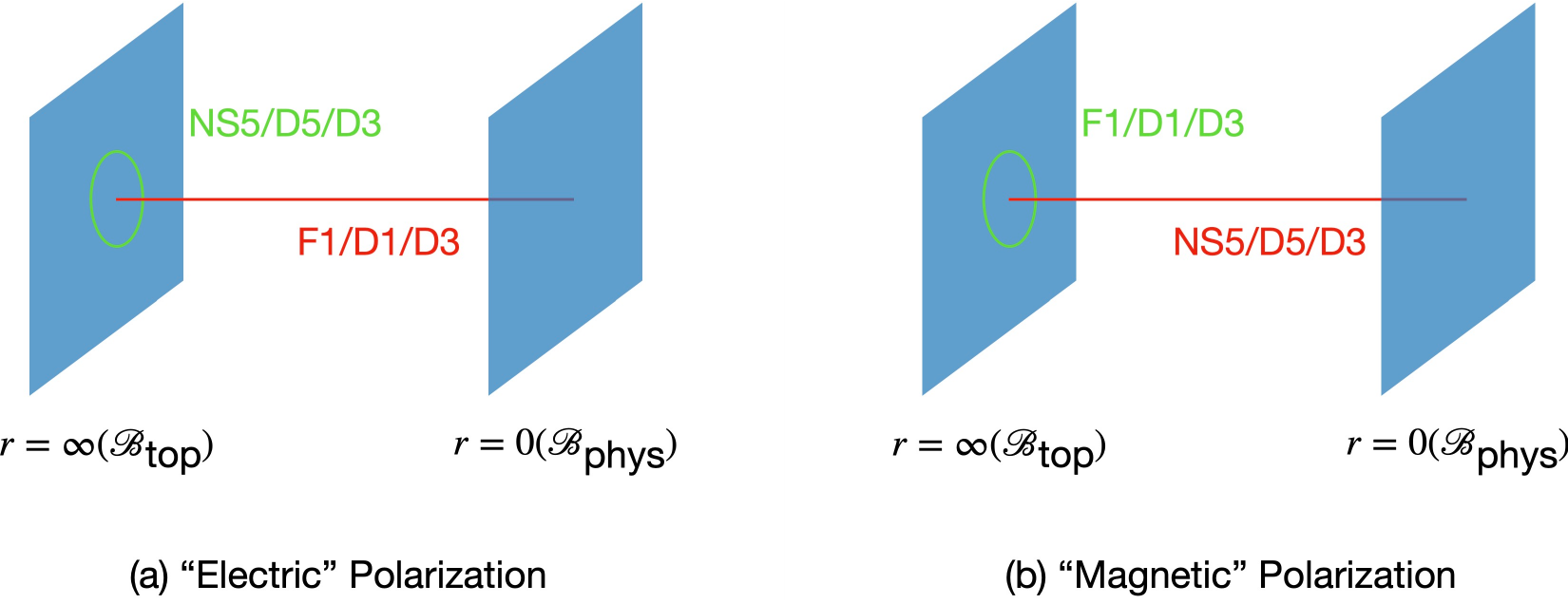}
    \caption{Brane origins of symmetry operators and charged objects under different polarizations, namely different topological boundary conditions. $r$ is the coordinate of the radial direction for the conical singularity of the Calabi-Yau 4-fold $Y$. At $r=0$, we have the singularity probed by D1-branes, while at $r=\infty$, boundary conditions for various SymTFT fields are picked. Symmetry operators are engineered from branes at $r=\infty$ (colored in green), while charged defects are built from branes terminating at $r=\infty$ (colored in red).}
    \label{fig: empolar}
\end{figure}

Now consider the case when the linking invariant (\ref{eq: linking a lines}) for the set $\{L_{(i)}^{\text{D3}} \}$ admits a Lagrangian subalgebra. Then, there are (at least) two gapped boundary conditions that can be picked: the ``electric'' one (\ref{eq: generic D condition for 3D SymTFT}) and the ``magnetic'' one (\ref{eq: generic magnetic D condition for 3D SymTFT}). The resulting invertible (\ref{eq: electric polarization symmetry}) and the non-invertible symmetries (\ref{eq: fusion rule of general non-invertible symmetries}) for the respective absolute QFTs can be seen as follows.
\begin{itemize}
    \item Under the ``electric'' condition (\ref{eq: generic D condition for 3D SymTFT}), F1-string and D1-string can terminate on the boundary due to the Dirichlet condition for $b_1^{(j)}$ and $c_1^{(j)}$ fields. This means F1-string and D1-string serve as local heavy objects charged under $G^{(b^{(j)})}=\Gamma_2^{(j)},~G^{(c^{(j)})}=\Gamma_2^{(j)}$ symmetries in (\ref{eq: electric polarization symmetry}) with their worldsheets wrapping on cone($\gamma_1^{(j)}$), stretching between the D1-branes worldvolume (where the 2D QFT is engineered) and infinity (where the asymptotic boundary $\partial Y$ is located). At the same time, their magnetic dual NS5-brane and D5-brane, serve as the topological operators at infinity along $\partial Y$. For the $L_{(k)}$ line from D3-brane, the non-invertible part from Dijkgraaf-Witten theories, i.e., the second line in (\ref{eq: general line operator from D3}), gets trivialized, while the invertible part survives as a topological line in the 2D QFT. This leads to the generator $D_{(k)}$ for the $G^{a^{(k)}}$ symmetry in the absolute QFT,
    \begin{equation}
        L^{\text{D3}}_{(k)}\rightarrow D_{(k)}\equiv \exp\left(2\pi i \int_{M_1} \sum_{i}\Lambda_{ki}a_1^{(i)} \right) .
    \end{equation}
    In total, the 2D QFT enjoys an invertible symmetry (\ref{eq: electric polarization symmetry}), which we reproduce below for ease of reading:
\begin{equation}
G^{(a^{(k)})}=\Gamma_4^{(k)},~G^{(b^{(j)})}=\Gamma_2^{(j)},~G^{(c^{(j)})}=\Gamma_2^{(j)},
\end{equation}
See Figure \ref{fig: empolar} (a) for a schematic illustration of the brane configuration under this ``electric" boundary condition.
    \item Under the ``magnetic'' condition (\ref{eq: generic magnetic D condition for 3D SymTFT}), F1-string and D1-string are still dynamical along the asymptotic boundary $\partial Y$ in string theory (or field-theoretically, gapped boundary of the SymTFT). This means F1-string and D1-string serve as topological line operators generating the quantum symmetries $G^{(\hat{b}_{(j)})}\times G^{(\hat{c}_{(j)})}$ in (\ref{eq: quantum symmetries for b and c}) via gauging the $G^{(b^{(j)})}\times G^{(c^{(j)})}$ symmetry. NS5-brane and D5-brane, ending at infinity along $\partial Y$, build the charged objects for this quantum symmetry. For the $L_{(k)}^{\text{D3}}$ line from D3-branes, both invertible and non-invertible parts are not trivialized. Therefore, in the resulting 2D QFT, one ends up with a non-invertible line defect $\mathcal{N}_{(k)}$ as well as two invertible defects $\eta_{b^{(j)}}$ and $\eta_{c^{(j)}}$
    \begin{equation}
      L^{\text{D3}}_{(k)}\rightarrow \mathcal{N}_{(k)},~ L^{(\text{F1})}_{(j)}\rightarrow \eta_{b^{(j)}},~ L^{(\text{D1})}_{(j)}\rightarrow \eta_{c^{(j)}},
    \end{equation}
    whose fusion rules are given by (\ref{eq: fusion rule of general non-invertible symmetries}), generating 
    \begin{equation}
        G=\text{TY}(\Gamma_2^{(j)}\times \Gamma_2^{(j)})~\text{fusion categorical symmetry.}
    \end{equation}
    See Figure \ref{fig: empolar} (b) for a schematic illustration of the brane configuration under this ``magnetic" boundary condition.
\end{itemize}
Note that the above discussion also reproduces (\ref{eq: line behavior under different conditions}), namely the invertible symmetry $G^{a^{(k)}}$ and its non-invertible symmetry promotion correspond to the same operator in the SymTFT bulk, whose stringy origin is a D3-brane.

The anomalies of these non-invertible symmetries are encoded back in the linking invariant (\ref{eq: linking a lines}). If elements in the set $\{ L_{(k)}^{\text{D3}} \}\subset \{ L_{(i)}^{\text{D3}} \}$ generating the $G^{a^{(k)}}$ and the non-invertible defects $\mathcal{N}_{(k)}$ always have non-trivial linking among them, then it is not possible to simultaneously condensing all lines in $\{L_{(k)}^{\text{D3}} \}$. This implies an anomaly via the obstruction to the gapped boundary condition as an obstruction to gauging the  $G^{a^{(k)}}$ symmetry or gauging the non-invertible symmetry generated by $\mathcal{N}_{(k)}$.

\subsubsection{Hanany-Witten Transition}

\begin{figure}[h]
    \centering
    \includegraphics[width=15cm]{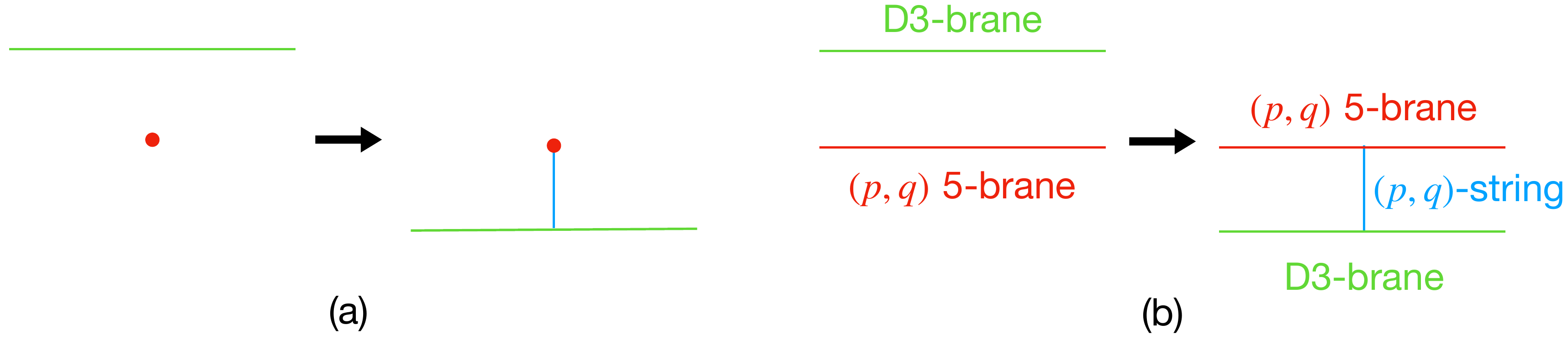}
    \caption{(a) Motion of the non-invertible symmetry operator passing through a local operator. A topological line is created, attaching to the non-invertible line and the local operator. (b) This non-trivial action is realized in string theory as the Hanany-Witten transition. The string theory origins for various field-theoretic objects are given by (\ref{eq: general line operator from D3}), (\ref{eq: general line operators for F1 and D1}), and (\ref{eq: charged objects from 5-branes}).}
    \label{fig: HWtransition}
\end{figure}

One of the typical properties of non-invertible symmetry operators is their behavior when passing through the charged objects. In 2D, moving a non-invertible line operator past a genuine local operator leaves behind another topological line attached to a defect operator and a T-shape junction. See Figure \ref{fig: HWtransition} (a).

This non-trivial action of non-invertible symmetries admits a string theory interpretation as the Hanany-Witten transition \cite{Hanany:1996ie}.
Consider the ``magnetic'' polarization (\ref{eq: generic magnetic D condition for 3D SymTFT}) where the associated 2D QFTs enjoy TY$(\Gamma_2^{(j)}\times \Gamma_2^{(j)})$ categorical symmetry. One can build local charge operators from 
\begin{equation}\label{eq: charged objects from 5-branes}
    \text{$(p,q)$ 5-brane wrapping on cone}(\gamma_{5(j)})
\end{equation}
where $p$ and $q$ label the D5-brane and NS5-brane charges, respectively. Recall that these wrapping branes are sources for the $\hat{b}$ and $\hat{c}$ fields in the SymTFT (\ref{eq: SymTFT for 4-fold}), implying their linkings with D1-strings and F1-strings, as discussed in (\ref{eq: F1-NS5 and D1-D5 pairings}). Therefore, the above $(p,q)$ 5-brane gives rise to the local operator carrying 
charge $(q,p)$ under the invertible part $G^{(\hat{b}_{(j)})}\times G^{(\hat{c}_{(j)})}$ within the TY$(\Gamma_2^{(j)}\times \Gamma_2^{(j)})$ categorical symmetry:
\begin{gather*}
    \text{$(p,q)$ 5-brane wrapping on cone}(\gamma_{5(j)})\\
    \updownarrow \\
    \text{$(q,p)$-charged local operator under $G^{(\hat{b}_{(j)})}\times G^{(\hat{c}_{(j)})}$}.
\end{gather*}

Now, consider a wrapped D3-brane moving through this $(p,q)$ 5-brane along the transverse direction of the 2D spacetime. A $(p,q)$-string\footnote{For the reader not familiar with the notation: A $(p,q)$-string is a type IIB S-duality covariant bound state carrying $p$ charges of F1-string and $q$ charges of D1-string.} will be created and attached between the D3-brane and the $(p,q)$ 5-brane. This created string is exactly the topological line operator (\ref{eq: general line operators for F1 and D1}) generating the $G^{(\hat{b}_{(j)})}\times G^{(\hat{c}_{(j)})}$ symmetry. This Hanany-Witten transition configuration exactly matches the non-trivial action of the non-invertible symmetry line on local charged objects. See Figure \ref{fig: HWtransition} (b). We remark that this Hananay-Witten origin for actions of non-invertible symmetries is ubiquitous for QFTs admitting string theory embeddings. See, e.g., \cite{Apruzzi:2022rei, Heckman:2022xgu, Yu:2023nyn, Apruzzi:2023uma}.

\section{Examples}

In this section, we discuss explicit examples of 2D QFTs engineered from Calabi-Yau 4-folds with D1-brane probes. The following three subsections correspond to the three facets of our general discussion in previous sections: intrinsic relativeness, non-invertible symmetries, and the manifestation of their anomalies as obstructions to gauging. For simplicity, we will only provide quiver diagrams but not the full J- and E-terms, since we are not making use of them.

\subsection{$Y^{p,k}(\mathbb{P}^2)$: When are they Intrinsically Relative?}

In Section 4.1.1, we obtained a sufficient condition for the intrinsic relativeness of 2D QFTs associated with Calabi-Yau 4-folds. We now apply this condition to an infinite class of Calabi-Yau 4-folds, namely cones over Sasaki-Einstein 7-manifold $Y=Y^{p,k}(\mathbb{P}^2)$. 

The toric data of this class of Calabi-Yau 4-folds is given by a convex polytope with $\Z^4$ coordinates \cite{Martelli:2008rt}
\begin{equation}
    p_1=(0,0,0,1), p_2=(0,0,p,1), p_3=(1,0,0,1), p_4=(0,1,0,1), p_5=(-1,-1,k,1).
\end{equation}
The associated 2D QFTs living on D1-brane probes have quiver gauge theory descriptions constructed in \cite{Franco:2022isw}, to which we refer the reader for more details.

The Sasaki-Einstein base space $Y^{p,k}(\mathbb{P}^2)$ gives rise to the asymptotic boundary $\partial Y$, with the cohomology classes \cite{Martelli:2008rt}
\begin{equation}\label{eq: coho class of Ypkp2}
    H^*(S^7/\Z_2; \Z)=\left\{\mathbb{Z},0,\Z\oplus \Z_{\text{gcd}(p,k)},0,\Gamma_4, \Z,\Z_{\text{gcd}(p,k)},\mathbb{Z} \right\},
\end{equation}
where $\Gamma_4=\Z^2/\langle (0,-3p+k),(k,p) \rangle$. Using the Smith normal form decomposition,\footnote{See the computation, e.g., https://en.wikipedia.org/wiki/Smith\_normal\_form} we can derive the general result (see also \cite{vanBeest:2022fss})
\begin{equation}
    \Gamma_4=\Z_{\text{gcd}(p,k)}\times\Z_{\frac{k(3p-k)}{\text{gcd}(p,k)}}.
\end{equation}
It is easy to see the second, fourth, and sixth cohomology classes are purely torsional: $\text{Tor}H^{(i)}(S^7/\Z_2;\Z_2)=H^{(i)}(S^7/\Z_2;\Z_2), i=2,4,6$. For each of these classes, there is one corresponding differential cohomology generator 
\begin{equation}
    \breve{t}_{i}\in \breve{H}^i(S^7/\Z_2).
\end{equation}

The expansion of differential cohomology uplifts of various IIB fluxes in this example is
\begin{equation}\label{eq: expansion of diff cochains c4z2}
\begin{split}
     \breve{F}_{6}=&\breve{a}_4\star \breve{t}_{2}+\breve{a}_2^{(1)}\star \breve{t}_{4(1)}+\breve{a}_2^{(2)}\star \breve{t}_{4(2)}\cdots,\\
     \breve{H}_{3}=&\breve{b}_{1}\star \breve{t}_{2}+\cdots,\\
    \breve{G}_{3}=&\breve{c}_1\star \breve{t}_{2}+\cdots,
\end{split}
\end{equation}
where we label the background fields for $\Z_{\text{gcd}(p,k)}$ and $\Z_{\frac{k(3p-k)}{\text{gcd}(p,k)}}$ by cochains $\breve{a}_1^{(1)}$ and $\breve{a}_1^{(2)}$, respectively. The corresponding SymTFT terms in (\ref{eq: SymTFT for 4-fold}) read
\begin{equation}
    2\pi \int_{M_3}\Lambda_{i_1i_2}a_1^{(i_1)}\cup \delta a_1^{(i_2)}.
\end{equation}

As we discussed in Section 2 and Section 4.1.1, a sufficient condition for intrinsic relativeness is when $|\Gamma_4|$ is not a perfect square, due to the absence of the Lagrangian subgroup of the defect group. For this class of theories, we arrive at the following statement:
\begin{equation*}
\boxed{
\text{\emph{The 2D QFT for $Y^{p,k}(\mathbb{P}^2)$ is intrinsically relative if $k(3p-k)$ is not a perfect square.}}
} 
\end{equation*}

One can perform a similar analysis for other classes of theories, e.g., orbifold singularities and $Y^{p,k}(\mathbb{P}^1\times \mathbb{P}^1)$, and obtain the sufficient condition for 2D QFTs being intrinsically relative.  In the following subsections, we will focus on Calabi-Yau 4-folds admitting absolute QFTs and their non-invertible symmetries, possibly with anomalies.

\subsection{$Y^{2,0}(\mathbb{P}^1\times \mathbb{P}^1)$: Rep$(D_4)$ Symmetry}

\begin{figure}[h]
    \centering
    \includegraphics[width=6cm]{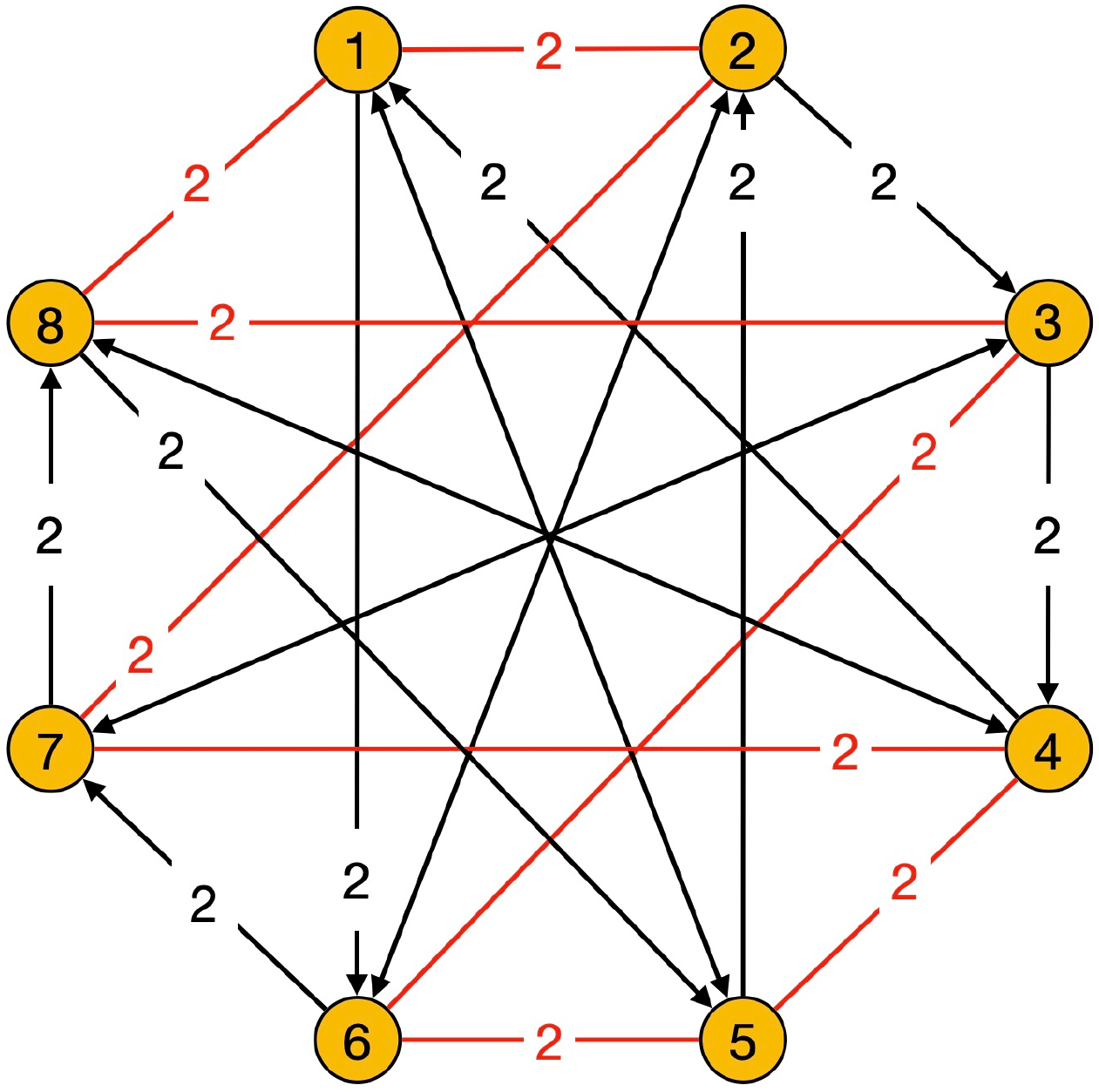}
    \caption{Quiver diagram for $Y^{2,0}(\mathbb{P}^1\times \mathbb{P}^1)$ probed by D1-branes}
    \label{fig: y20quiver}
\end{figure}

Let us now consider the Calabi-Yau 4-fold given by the cone over the Sasaki-Einstein 7-manifold $Y^{2,0}(\mathbb{P}^1\times \mathbb{P}^1)$, which belongs to an infinite class of geometries known as $Y^{p,k}(B_4)$ \cite{Martelli:2008rt}, where $B_4$ denotes a four-dimensional base space that is either $\mathbb{P}^1\times \mathbb{P}^1$ or $\mathbb{P}^2$. The 2D QFTs associated with this general class of Calabi-Yau 4-folds were systematically constructed and studied in \cite{Franco:2022isw}. The toric data for $Y^{2,0}(\mathbb{P}^1\times \mathbb{P}^1)$ is given by the polytope with the following vertex coordinates in $\Z^4$ \cite{Franco:2022isw}
\begin{equation}
\begin{split}
    &p_1=(1,1,0,0), p_2=(1,0,1,0), p_3=(1,-1,0,0), p_4=(1,0,-1,0), p_5=(1,0,0,2),\\
    &s=(1,0,0,1), q=(1,0,0,0)
\end{split}
\end{equation}
The associated 2D QFT living on D1-brane probes enjoys a quiver gauge theory description, with quiver diagram given in Figure \ref{fig: y20quiver} \footnote{For a given Calabi-Yau 4-fold, the associated 2D gauge theories are not unique. Here we pick one of the phases for the $Y^{(2,0)}(\mathbb{P}^1\times \mathbb{P}^1)$ found in \cite{Franco:2022isw}. Different 2D QFTs for a given geometry are connected via the $\mathcal{N}=(0,2)$ triality \cite{Gadde:2013lxa}.}\footnote{It is natural to ask how the non-invertible symmetries interplay with triality. As happens for ordinary global symmetries, we expect the dual QFTs connected by triality to enjoy the same non-invertible global symmetries. This can be understood since they share the same asymptotic boundary geometry in the string theory background. Despite this general expectation, it would still be interesting to implement non-invertible symmetries at the level of quivers and see how they interplay with the quiver mutations (i.e., field-theory trialities).}. We refer the reader to \cite{Franco:2016fxm, Franco:2022isw} additional details on this theory.

The SymTFT for finite symmetries of this 2D QFT was constructed in \cite{Yu:2023nyn}. Here, we revisit the result and present some further discussion. The cohomology classes of $Y^{2,0}(\mathbb{P}^1\times \mathbb{P}^1)$ read
\begin{equation}
    H^*(Y^{2,0}(\mathbb{P}^1\times \mathbb{P}^1); \Z)=\left\{ \mathbb{Z},0,\mathbb{Z}^2\oplus \mathbb{Z}_2, 0, \mathbb{Z}\oplus \mathbb{Z}_2\oplus \mathbb{Z}_2, \mathbb{Z}^2, \mathbb{Z}_2, \mathbb{Z} \right\}
\end{equation}
The differential cohomology generators for torsional parts read
\begin{equation}
\begin{split}
    &\breve{t}_2\in \breve{H}^2(Y^{2,0}(\mathbb{P}^1\times \mathbb{P}^1)), ~\breve{t}_6\in \breve{H}^6(Y^{2,0}(\mathbb{P}^1\times \mathbb{P}^1))\\
    &\breve{t}_{4(1)}\in \breve{H}^4(Y^{2,0}(\mathbb{P}^1\times \mathbb{P}^1)), ~\breve{t}_{4(2)}\in \breve{H}^4(Y^{2,0}(\mathbb{P}^1\times \mathbb{P}^1)), 
\end{split}
\end{equation}
The expansion of uplifts of various IIB fluxes is 
\begin{equation}\label{eq: expansion of diff cochains y20}
\begin{split}
     \breve{F}_{6}=&\breve{a}_4\star \breve{t}_{2}+\breve{a}_{2}\star \breve{t}_{4(1)}+\breve{\hat{a}}_{2}\star \breve{t}_{4(2)}\cdots,\\
     \breve{H}_{3}=&\breve{b}_{1}\star \breve{t}_{2}+\cdots,\\
    \breve{G}_{3}=&\breve{c}_1\star \breve{t}_{2}+\cdots.
\end{split}
\end{equation}
The linking numbers between various cohomology generators are given by \cite{Yu:2023nyn}
\begin{equation}\label{eq: linking numbers for y20}
\begin{split}
    &\Lambda_{21}=\Lambda_{12}=\int \breve{t}_{4(1)}\star \breve{t}_{4(2)}~\text{mod 1}=\frac{1}{2},\\ ~ &\Delta_{i}= -\int\breve{t}_{4(1)}\star \breve{t}_{2} \star \breve{t}_{2}~\text{mod 1}=\frac{1}{2},\\
    &\Omega=\int \breve{t}_2\star \breve{t}_6=\frac{1}{2}.
\end{split}
\end{equation}
Substituting these linking numbers in the general form of the SymTFT (\ref{eq: SymTFT for 4-fold}), one obtains the SymTFT for the $Y^{2,0}(\mathbb{P}^1\times \mathbb{P}^1)$ theory \cite{Yu:2023nyn}
\begin{equation}
    S_3=\frac{2\pi}{2}\int_{M_3} a_1 \cup \delta \hat{a}_1+b_1\cup \delta \hat{b}_1+c_1\cup \delta \hat{c}_1+a_1\cup b_1\cup c_1
\end{equation}
where all fields are $\Z_2$ cochains, whose differential cochain counterparts are in (\ref{eq: expansion of diff cochains y20}) under obvious notations.

The topological line operators in this SymTFT can be computed from brane actions explicitly, following the steps in Section 4.2. Substituting the linking numbers (\ref{eq: linking numbers for y20}) into  the general results for wrapped D3-brane (\ref{eq: general line operator from D3}), F1-string and D1-string (\ref{eq: general line operators for F1 and D1}), one obtains
\begin{equation}\label{eq: D3, F1, D1 lines in y20}
\begin{split}
    &L^{D3}_{(1)}=\exp \left(\pi i \int_{M_1}\hat{a}_1 \right) \times \int \D \hat{\phi}_0 \D \phi_0\exp \left(\pi i \int_{M_1} \hat{\phi}_0\cup \delta \phi_{0}+c_1\cup \phi_0-b_1\cup \hat{\phi}_0 \right),\\
    &L^{D3}_{(2)}=\exp \left(\pi i \int_{M_1}a_1 \right),~L^{\text{F1}}=\exp \left( \pi i\int_{M_1} b_1\right),~L^{\text{D1}}=\exp \left( \pi i\int_{M_1} c_1\right).
\end{split} 
\end{equation}
In addition to the four line operators above, it is also possible to derive other two operators from NS5-brane and D5-brane wrapping on the torsional 5-cycle $\gamma_5$, following the same steps in Section 4.2,
\begin{equation}\label{eq: NS5 and D5 lines in y20}
\begin{split}
    L^{\text{NS5}}=\exp \left(\pi i \int_{M_1}\hat{b}_1 \right) \times \int \D \hat{\phi}_0 \D \phi_0\exp \left(\pi i \int_{M_1} \hat{\phi}_0\cup \delta \phi_{0}+c_1\cup \phi_0-a_1\cup \hat{\phi}_0 \right),\\
    L^{\text{D5}}=\exp \left(\pi i \int_{M_1}\hat{c}_1 \right) \times \int \D \hat{\phi}_0 \D \phi_0\exp \left(\pi i \int_{M_1} \hat{\phi}_0\cup \delta \phi_{0}+a_1\cup \phi_0-b_1\cup \hat{\phi}_0 \right),
\end{split}
\end{equation}

\begin{figure}[h]
    \centering
    \includegraphics[width=15cm]{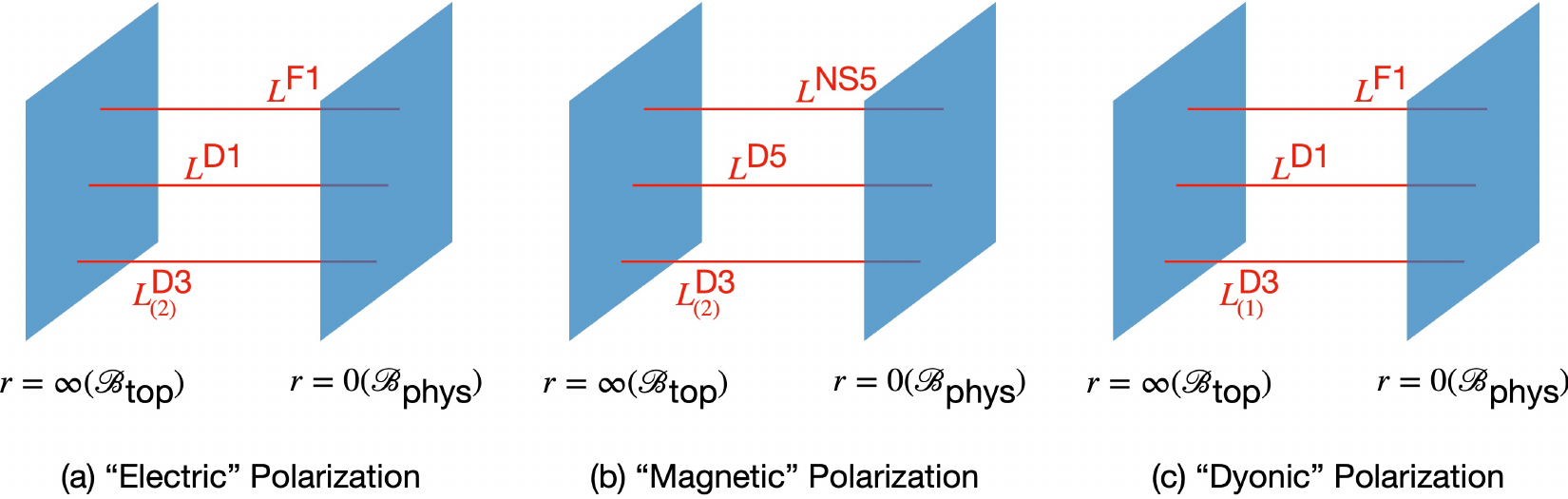}
    \caption{Brane origins for local operators in 2D QFTs for $Y^{(2,0)}(\mathbb{P}^1\times \mathbb{P}^1)$ under various boundary conditions at infinity. The global symmetries for these three polarizations are discussed in the main text as (a) Anomalous $(\Z_2)^3$ symmetry, (b) Rep$(D_4)$ symmetry, and (c) $D_4$ symmetry.}
    \label{fig: polarsy20}
\end{figure}

\paragraph{``Electric'' boundary condition for anomalous $(\Z_2)^3$ symmetry.}
The generic ``electric'' boundary condition (\ref{eq: generic D condition for 3D SymTFT}) in this example is
\begin{equation}
a_1,b_1,c_1~\text{Dirichlet},~\hat{a}_1,\hat{b}_1,\hat{c}_1~\text{Neumann}.
\end{equation}
Under this boundary condition, on the one hand, the three invertible line operators in the second line of (\ref{eq: D3, F1, D1 lines in y20}) are trivialized along the gapped boundary. This, in turn, leads to the fact the D3-brane on $\gamma_{3}^{(2)}$, F1-string and D1-string on $\gamma_1$ can end ``at infinity'', serving as the heavy charged objects. See Figure \ref{fig: polarsy20} (a) for a schematic depiction. On the other hand, the three non-invertible lines $L^{\text{D3}}_{(1)}, L^{\text{NS5}}$ and $L^{\text{D5}}$ are reduced to their invertible parts, generating the three $\Z_2$ symmetries 
\begin{equation}
    G^{(0)}=\mathbb{Z}_2^{(a)}\times \mathbb{Z}_2^{(b)}\times \mathbb{Z}_2^{(c)},
\end{equation}
aligning with the general form in (\ref{eq: electric polarization symmetry}) for the resulting absolute QFT. This $(\Z_2)^3$ symmetry has an anomaly given by the following 3D invertible TFT
\begin{equation}
    \exp \left( \pi i \int_{M_3}A\cup B\cup C \right),
\end{equation}
where $A, B$, and $C$ are background profiles for the $a_1, b_1$ and $c_1$ under Dirichlet gapped boundary conditions.\footnote{This anomaly is usually referred to as a type III anomaly, especially in the condensed-matter literature. See, e.g., \cite{deWildPropitius:1995cf}.}

\paragraph{``Magnetic'' boundary condition for Rep($D_4$) symmetry.} The generic ``magnetic'' boundary condition (\ref{eq: generic magnetic D condition for 3D SymTFT}) via gauging the $\Z_2^{(b)}\times \Z_2^{(c)}$ symmetry is
\begin{equation}\label{eq: magnetic condition for y20}
a_1,\hat{b}_1,\hat{c}_1~\text{Dirichlet},~\hat{a}_1,b_1,c_1~\text{Neumann}.
\end{equation}
Under this boundary condition, on the one hand, line operators $L_{(2)}^{\text{D3}}, L^{\text{NS5}}$ and $L^{\text{D5}}$ trivialize along the gapped boundary, on which they terminate and serve as charged defects for the 2D QFT. See Figure \ref{fig: polarsy20} (b) for an illustration. Interestingly, the non-invertible part of $L^{\text{D3}}_{(1)}$ survives in this case. Together with the invertible topological lines $L^{\text{F1}}$ and $L^{\text{D1}}$, this generates a non-invertible symmetry with TY$(\Z_2\times \Z_2)$ fusion rules
\begin{equation}\label{eq: fusion rules for y20}
\begin{split}
    &L^{\text{D3}}_{(1)}\otimes L^{\text{D3}}_{(1)}=1\oplus L^{\text{F1}}\oplus L^{\text{D1}}\oplus L^{\text{F1}}L^{\text{D1}},\\
    &L^{\text{D3}}_{(1)}\otimes L^{\text{F1}}=L^{\text{F1}}\otimes  L^{\text{D3}}_{(1)}=L^{\text{D3}}_{(1)},\\
      &L^{\text{D3}}_{(1)}\otimes L^{\text{D1}}=L^{\text{D1}}\otimes  L^{\text{D3}}_{(1)}=L^{\text{D3}}_{(1)},\\
      &L^{\text{F1}}\otimes  L^{\text{F1}}= L^{\text{D1}}\otimes  L^{\text{F1}}=1.
\end{split}
\end{equation}
The fusion rules, in general, do not fully fix the fusion category. For the TY$(\Z_2\times \Z_2)$ non-invertible symmetry, there are four fusion categories that obey the same fusion rules, three of which are non-anomalous, namely Rep$(D_4)$\footnote{We use the notation $D_n$ for order-$2n$ dihedral group. $D_4$ group in our notation has order 8, which sometimes is referred to as $D_8$ in other literature.}, Rep$(Q_8)$ and Rep$(\mathcal{H}_8)$\footnote{$\mathcal{H}_8$ is the eight-dimensional Kac-Paljutkin Hopf algebra \cite{kac1966finite}.} (see, e.g., \cite{Bhardwaj:2017xup, Thorngren:2021yso, Perez-Lona:2023djo, Diatlyk:2023fwf}). In order to identify the non-invertible symmetry with the above fusion rules, we can consider gauging the full non-invertible symmetry, which corresponds to the ``dyonic'' gapped boundary condition below for the SymTFT.

\paragraph{``Dyonic'' boundary condition for $D_4$ symmetry.}
Starting with the ``electric'' boundary condition with $(\Z_2)^3$ symmetry, we can consider gauging the $\Z_2^{(a)}$ symmetry, and end up with the following ``dyonic'' boundary condition
\begin{equation}
     \hat{a}_1,b_1,c_1~\text{Dirichlet}, ~a_1, \hat{b}_1, \hat{c}_1~\text{Neumann}.
\end{equation}
Under this condition, there are three $\Z_2$ symmetries $\Z_2^{(\hat{a})}$, $\Z_2^{(b)}$ and $\Z_2^{(c)}$ with respective background field profiles $\hat{A}, B$ and $C$ under Dirichlet boundary conditions. See Figure \ref{fig: polarsy20} (c) for brane origins of charged defects under this polarization. However, the genuine global symmetry here is not a direct product of the three $\Z_2$ factors, due to the extra condition
\begin{equation}
    \delta \hat{A}=BC
\end{equation}
inherited from the equation of motion of $a$ in the SymTFT.
 According to \cite{deWildPropitius:1995cf} (see also \cite{Tachikawa:2017gyf, Bhardwaj:2017xup, Wang:2014oya, Bhardwaj:2023kri}), the above equation leads to a group extension $\Z_2^{(\hat{a})}\rtimes (\Z_2^{(b)}\times \Z_2^{(c)})=D_4$\footnote{We thank Yunqin Zheng for pointing out relevant references. We also thank Jonathan J. Heckman, Max H\"{u}bner and Hao Y. Zhang for discussions on this point.}. 
 
 The ``magnetic'' boundary condition (\ref{eq: magnetic condition for y20}) can then be derived from gauging the whole $D_4$ symmetry. Note that the non-invertible topological line $L^{\text{D3}}_{(1)}$ generating the Rep$(D_4)$ symmetry under ``magenetic'' condition now ends at infinity and becomes charged defects under ``dyonic'' condition. See Figure \ref{fig: polarsy20} (c) for an illustration. This identifies the non-invertible symmetry with TY$(\Z_2\times \Z_2)$ fusion rules (\ref{eq: fusion rules for y20}) as the Rep$(D_4)$ categorical symmetry \cite{Bhardwaj:2017xup} and aligns with the fact that the Rep$(D_4)$ symmetry is non-anomalous and gaugeable with $D_4$ quantum symmetry. 

\begin{figure}[h]
     \centering
     \includegraphics[width=15cm]{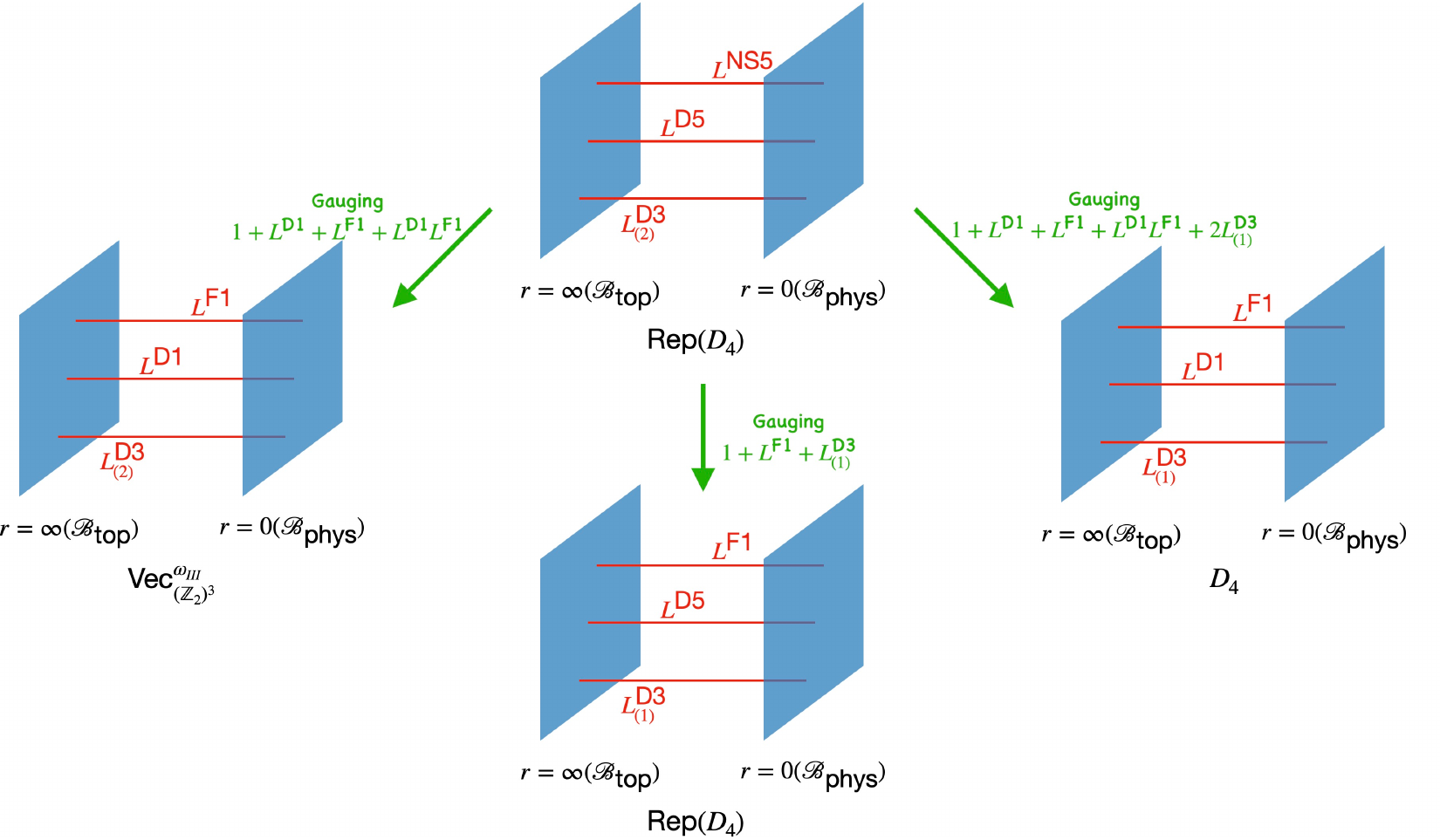}
         \caption{Some gauging manipulations for Rep$(D_4)$ symmetry and their top-down SymTFT picture. The ``electric'' and ``dyonic'' boundary conditions can be realized via gauging the $\mathbb{Z}_2\times \mathbb{Z}_2$ subcategory and gauging the whole Rep$(D_4)$, respectively. The resulting quantum symmetry for the ``electric'' polarization is denoted by Vec$_{(\mathbb{Z}_2)^3}^{\omega_{\text{III}}}$, meaning a $(\mathbb{Z}_2)^3$ symmetry with type III anomaly \cite{deWildPropitius:1995cf}.}
     \label{fig: groupoid}
 \end{figure}

 We conclude this subsection by remarking that this top-down SymTFT picture matches the generalized gauging of Rep$(D_4)$ symmetry discussed in \cite{Perez-Lona:2023djo,Diatlyk:2023fwf}. Namely, starting with the ``magnetic'' boundary condition for Rep$(D_4)$ symmetry, topological manipulations change this boundary condition to others, realizing different gauging choices of the Rep$(D_4)$ symmetry. A portion of the generalized gauging for Rep($D_4$) and its top-down SymTFT configuration is shown in Figure \ref{fig: groupoid}.

\subsection{$\mathbb{C}^4/\mathbb{Z}_4$: Anomalies of the Non-invertible Symmetry}

\begin{figure}[h]
    \centering
    \includegraphics[width=4.6cm]{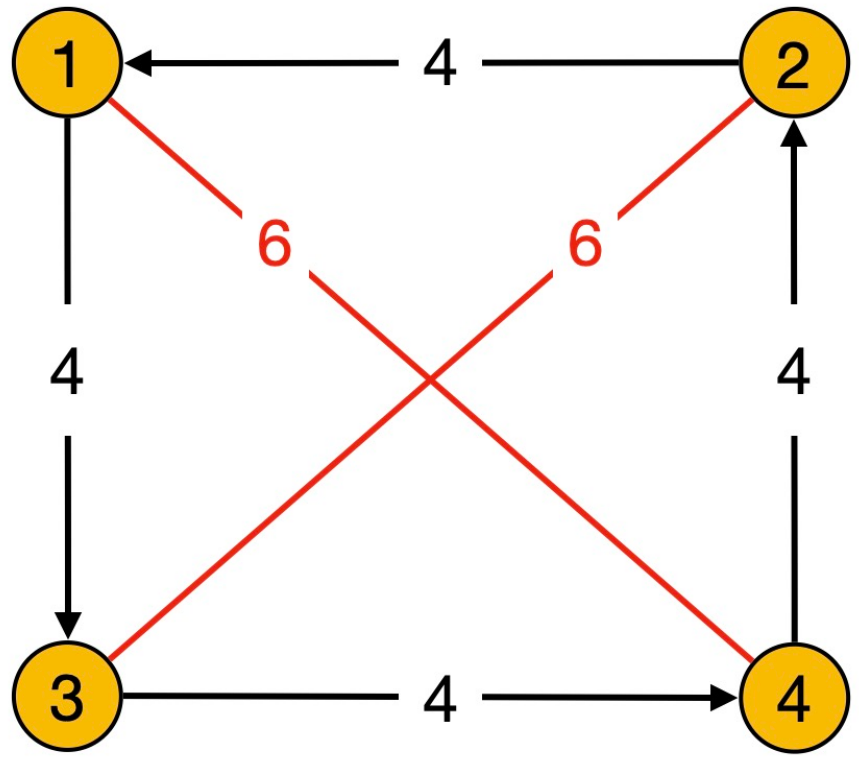}
    \caption{Quiver diagram for $\C^4/\Z_4$ probed by D1-branes.}
    \label{fig: c4z4 quiver}
\end{figure}

The local Calabi-Yau 4-fold $\C^4/\Z_4$ is defined by the following orbifold action
\begin{equation}
    (z_1,z_2,z_3,z_4)\rightarrow (e^{\pi i/2}z_1,e^{\pi i/2}z_2,e^{\pi i/2}z_3,e^{\pi i/2}z_4).
\end{equation}
The toric data of $\C^4/\Z_4$ is given by the polytope whose vertices have the following $\Z^4$ coordinates
\begin{equation}
    p_1=(1,1,0,0), p_2=(1,0,1,0), p_3=(1,0,0,1), p_4=(1,-1,-1,-1), q=(1,0,0,0).
\end{equation}
The associated 2D QFT living on D1-brane probes has a quiver gauge theory description. Figure \ref{fig: c4z4 quiver} shows the corresponding quiver diagram.
We refer the reader to \cite{Franco:2015tna} for detailed information on this theory.

The asymptotic boundary in this case is simply the seven-dimensional lens space $S^7/\Z_4$, whose cohomology classes read
\begin{equation}
    H^*(S^7/\Z_2; \Z)=\left\{\mathbb{Z},0,\Z_4,0,\Z_4, 0,\Z_4,\mathbb{Z} \right\}.
\end{equation}
The expansion of differential uplifts of various IIB fluxes is the same as in $\C^4/\Z_4$
\begin{equation}\label{eq: expansion of diff cochains c4z4}
\begin{split}
     \breve{F}_{6}=&\breve{a}_4\star \breve{t}_{2}+\breve{a}_2\star \breve{t}_{4}+\cdots,\\
     \breve{H}_{3}=&\breve{b}_{1}\star \breve{t}_{2}+\cdots,\\
    \breve{G}_{3}=&\breve{c}_1\star \breve{t}_{2}+\cdots,
\end{split}
\end{equation}
The linking numbers between various cohomology generators can be computed as (see also \cite{vanBeest:2022fss})
\begin{equation}\label{eq: linking numbers for c4z4}
\begin{split}
    &\Lambda_{21}=\Lambda_{12}=\int \breve{t}_{4}\star \breve{t}_{4}~\text{mod 1}=\frac{1}{4},\\ ~ &\Delta_{i}= -\int\breve{t}_{4}\star \breve{t}_{2} \star \breve{t}_{2}~\text{mod 1}=\frac{1}{4},\\
    &\Omega=\int \breve{t}_2\star \breve{t}_6=\frac{1}{4}.
\end{split}
\end{equation}

Substituting these linking numbers in the general form of the SymTFT (\ref{eq: SymTFT for 4-fold}), one obtains the SymTFT for the $\C^4/\Z_4$ theory
\begin{equation}
    S_3=\frac{2\pi}{4}\int_{M_3} \frac{1}{2}a_1 \cup \delta a_1+b_1 \cup \delta \hat{b}_1+c_1 \cup \delta \hat{c}_1-a_1\cup b_1 \cup c_1 ,
\end{equation}
where all gauge fields are $\Z_4$-valued cochains, whose differential cochain counterparts are in (\ref{eq: expansion of diff cochains c4z4}) under obvious notations. The defect group is given by
\begin{equation}
    \mathbb{D}=\Z_4^{(a)}\times \Z_4^{(b)}\times \Z_4^{(\hat{b})} \times \Z_4^{(c)} \times \Z_4^{(\hat{c})},
\end{equation}
with Dirac pairing matrix extracted from the coefficients of the single-derivative terms in the SymTFT
\begin{equation}\label{eq: dirac pairing for c4z4}
    \begin{pmatrix}
        \textcolor{red}{\frac{1}{4}}& 0 & 0 & 0 & 0 \\
        0 & 0 & \frac{1}{4} & 0 & 0 \\
        0& \frac{1}{4} & 0 & 0 & 0 \\
        0& 0 & 0 & 0 & \frac{1}{4} \\
        0& 0 & 0 & \frac{1}{4} & 0 \\
    \end{pmatrix}.
\end{equation}

The topological line operators from D3-branes, F1-strings and D1-strings can be computed from the general form in (\ref{eq: general line operator from D3}) and (\ref{eq: general line operators for F1 and D1}) using the linking number (\ref{eq: linking numbers for c4z4}):
\begin{equation}\label{eq: lines from branes c4z4}
\begin{split}
     &L^{D3}=\exp \left(\frac{\pi i}{2} \int_{M_1}\hat{a}_1 \right) \times \int \D \hat{\phi}_0 \D \phi_0\exp \left(\frac{\pi i}{2} \int_{M_1} \hat{\phi}_0\cup \delta \phi_{0}+c_1\cup \phi_0-b_1\cup \hat{\phi}_0 \right),\\
&L^{\text{F1}}=\exp \left(\frac{ \pi i}{2}\int_{M_1} b_1\right),~L^{\text{D1}}=\exp \left(\frac{ \pi i}{2}\int_{M_1} c_1\right).     
\end{split}
\end{equation}
One can compute similarly line operators from NS5-branes and D5-branes wrapping on the torsional 5-cycle. Their concrete expression is not important for our following discussion, so we will just simply donote them as $L^{\text{NS5}}$ and $L^{\text{D5}}$.

\begin{figure}[h]
    \centering
    \includegraphics[width=12cm]{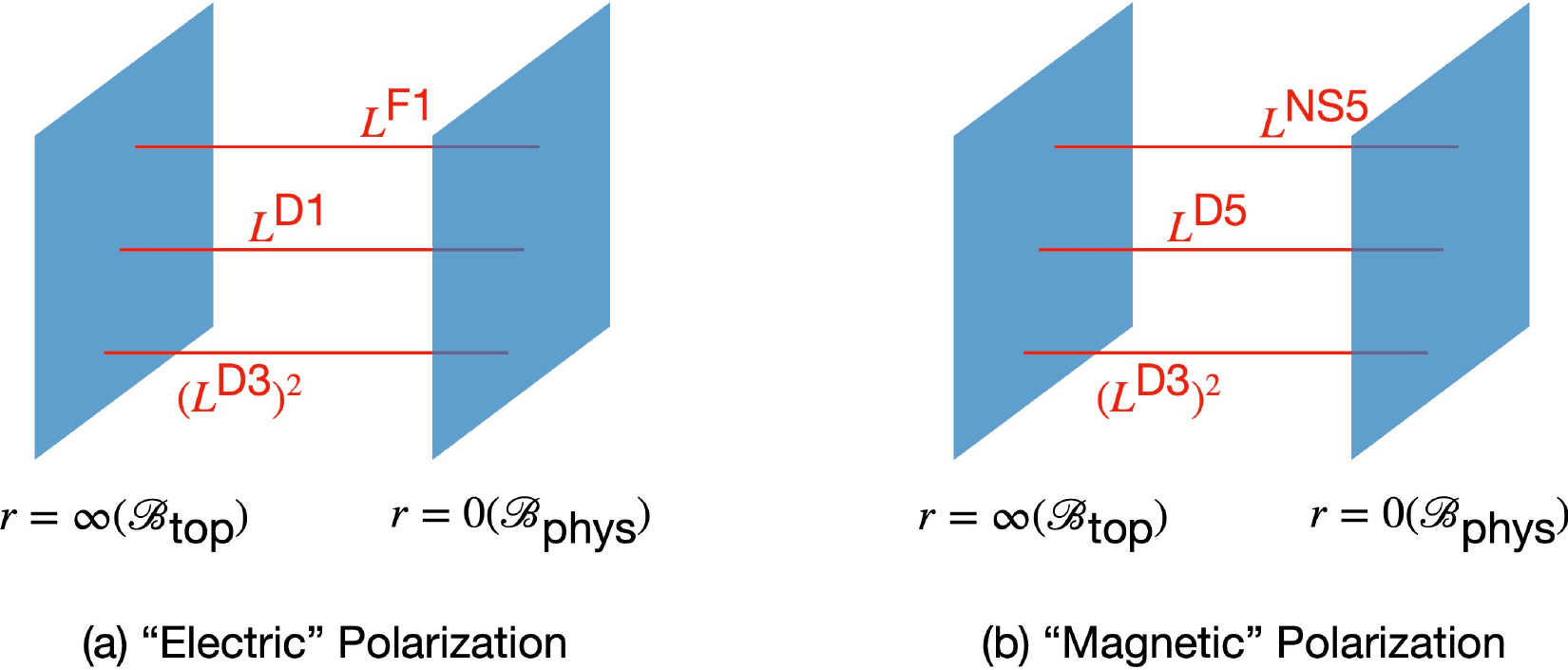}
    \caption{Brane origins for local operators in 2D QFTs for $\C^4/\Z_4$ under various boundary conditions at infinity. $(L^{(D3)})^2$ is engineered from a collection of two D3-branes. The global symmetries for these three polarizations are discussed in the main text as (a) Anomalous $\Z_2\times (\Z_4)^2$ symmetry, and (b) Anomalous TY($\Z_4\times \Z_4$) non-invertible symmetry.}
    \label{fig: polarsc4z4}
\end{figure}

\paragraph{``Electric'' boundary condition for $\Z_2\times (\Z_4)^2$ symmetry.} First, consider the polarization with the following Lagrangian subgroup 
\begin{equation}
    L=\Z_2^{(a)}\times \Z_4^{(\hat{b})} \times \Z_4^{(\hat{c})}.
\end{equation}
The corresponding gapped boundary condition of the SymTFT for $b_1$ and $c_1$ are standard Dirichlet boundary conditions, while for $a_1$ it is a bit special:
\begin{equation}
    a_1|_{\partial M_3}=A_1 .
\end{equation}
Similarly to (\ref{eq: boundary condition of U(1)4}), the above condition constrains the $\Z_4$-valued $a_1$  to a $\Z_2$-valued background gauge field profile $A_1$, implying the quotient by $\Z_2^{(a)}\subset \Z_4^{(a)}$ Lagrangian subgroup. In the SymTFT language, this is to say $L^{\text{D3}}$ does not compose any Lagrangian subalgebra, namely there is no gapped boundary for it to ends. However, condensing the subalgebra $1\oplus (L^{\text{D3}})^2$ is admitted, and it define a topological boundary on which $(L^{\text{D3}})^2$ can end. From the string theory perspective, this means the boundary condition at infinity does not allow a single D3-brane to end, but a collection of two D3-branes can terminate at infinity, giving rise to the charged defect. See Figure \ref{fig: polarsc4z4} (a) for an illustration.

The resulting absolute QFT enjoys an invertible $(\Z_2)^3$ global symmetry, as discussed in the general form (\ref{eq: electric polarization symmetry}), reading
\begin{equation}
    G=L^\vee=\Z_2^\vee \times \Z_4^{(b)} \times \Z_4^{(c)},
\end{equation}
with $\Z_2^\vee=\Z_4^{(a)}/\Z_2^{(a)}$ due to  the exact sequence
\begin{equation}
    1\rightarrow \Z_2^{(a)} \rightarrow \Z_4^{(a)} \rightarrow \Z_2^{\vee} \rightarrow 1.
\end{equation}
Furthermore, this $\Z_2\times (\Z_4)^2$ symmetry suffers from anomalies captured by the following invertible 3D TFT 
\begin{equation}\label{eq: anomaly tft for c4z4}
    \frac{2\pi}{4}\int_{M_3} A_1\cup \delta A_1-A_1\cup B_1 \cup C_1.
\end{equation}
where $B_1$ and $C_1$ are Dirichlet boundary profiles for $b_1$ and $c_1$ fields. In the above invertible TFT, the first term shows a self 't Hooft anomaly for the $\Z_2^\vee$ symmetry, while the second cubic term captures a mixed anomaly between $\Z_2^\vee$,  $\Z_4^{(b)}$ and $\Z_4^{(c)}$.

\paragraph{``Magnetic'' boundary condition for anomalous TY$(\Z_4\times Z_4)$ symmetry.}
Based on our discussion in Section 4, one can gauge the $\Z_4^{(b)}\times \Z_4^{(c)}$ symmetry and end up with the following Lagrangian subgroup 
\begin{equation}
    L=\Z_2^{(a)}\times \Z_4^{(b)} \times \Z_4^{(c)}.
\end{equation}
The corresponding gapped boundary condition of the SymTFT for $a_1$ is not changed, but now $\hat{b}_1$ and $\hat{c}_1$ get Dirichlet boundary conditions. In order to read the global symmetry for the resulting absolute QFT, notice that
the associated quotient for the defect group is 
\begin{equation}
    L^\vee=\mathbb{D}/L\cong \Z_2^{\vee}\times \Z_4^{(\hat{b})} \times \Z_4^{(\hat{c})}.
\end{equation}
However, this is not the genuine global symmetry. Instead, with the presence of the cubic anomaly term $(2\pi/4)\int_{M_3}A_1\cup B_1 \cup C_1$ in (\ref{eq: anomaly tft for c4z4}) under the ``electric" polarization, the $\Z_2^{\vee}$ part is not a direct product with the other two factors but is promoted to a non-trivial $\Z_2$ extension.\footnote{Conceptually, this $\Z_2$ extension is due to the fact that the TY$(G)$ non-invertible symmetry can be realized from the self-duality under gauging $G$, with $G$ a group.} The resulting symmetry reads 
\begin{equation}
    \text{TY$(\Z_4^{(\hat{b})}\times \Z_4^{(\hat{c})})$ fusion categorical symmetry}.
\end{equation}
The fusion rules, falling in the general form (\ref{eq: fusion rule of general non-invertible symmetries}), for this non-invertible symmetry can be computed explicitly from the topological line operators engineered from D3-brane, F1-string and D1-string in (\ref{eq: lines from branes c4z4}).

According to our discussion in Sections 2.3.2 and 4.2.1, it is now straightforward to see that this non-invertible symmetry is anomalous. Let us assume this symmetry is gaugeable. From the defect group point of view, this is to say there is a polarization for which the Lagrangian subgroup is given by the $L^\vee$ uplift. However, due to the $\frac{1}{4}$ factor in red in (\ref{eq: dirac pairing for c4z4}), $L^\vee$ cannot be embedded back in the $\mathbb{D}$ as a Lagrangian subgroup. This means the assumed polarization of the defect group obtained from gauging the TY$(\Z_4\times \Z_4)$ symmetry, in fact, does not exist, implying this non-invertible symmetry is anomalous. From the SymTFT and its string theory origin point of view, this translates to the fact that only a collection of two D3-branes wrapping on torsional 3-cycles are allowed to end at infinity, due to the self-linking property of D3-branes. This prevents the existence of a ``dyonic'' boundary condition (named after the $Y^{(2,0)}(\mathbb{P}^1\times \mathbb{P}^1)$ example in Section 5.2), where other lines built from D3-branes can terminate. In other words, the boundary condition after an attempt of gauging TY$(\Z_4\times \Z_4)$ symmetry is obstructed, implying an anomaly of this non-invertible symmetry.

\acknowledgments

We thank I. Garcia Etxebarria, J. J. Heckman, M. H\"{u}bner, H. T. Lam, Y.-H. Lin, S. Schafer-Nameki, S.-H. Shao, E. Sharpe, E. Torres, Y. Wang, H. Y. Zhang, and Y. Zheng for helpful discussions. XY thanks J. J. Heckman, M. H\"{u}bner, C. Lawrie, E. Torres, and H. Y. Zhang for their collaboration on related projects. We thank the 2023 Simons Summer Workshop for hospitality during part of this work. XY thanks the 2023 NYU Satellite Workshop on Global Categorical Symmetries, the 2023 Annual Meeting of the Simons Collaboration on Global Categorical Symmetries, and the 2024 Spring Southeastern Regional
Mathematical String Theory Meeting for hospitality during part of this work. SF is supported by the U.S. National Science Foundation grants PHY-2112729 and DMS-1854179. XY is supported by NSF grant PHY-2014086.

\appendix

\bibliographystyle{utphys}
\bibliography{catsbbm}
\end{document}